%% file: main.tex
\documentclass[conference,compsoc]{IEEEtran}

\input{preamble}

\begin{document}

\title{SoK: Blockchain Decentralization}

%\author{anynomous author(s)}

\author{

\IEEEauthorblockN{
Luyao Zhang$^{1}$\symbolfootnote{*}\textsuperscript{\textsection}, 
Xinshi Ma$^{2}$\textsuperscript{\textsection}, 
Yulin Liu$^{3}$\symbolfootnote{*}\textsuperscript{\textsection}
}

\IEEEauthorblockA{
$^{1}$ Data Science Research Center and Social Science Division, Duke Kunshan University, Jiangsu, China}%
\IEEEauthorblockA{
$^{2}$ Duke University, Durham, United States}
\IEEEauthorblockA{
$^{3}$ Bochsler Consulting, Zurich, Switzerland }
}

\maketitle
\iffalse
\footnotetext[1]{joint corresponding authors: Luyao Zhang (email: lz183@duke.edu, address:  Duke Kunshan University, No.8 Duke Ave., Kunshan, Jiangsu, China, 215316); Yulin Liu (email:yulin@bochslerfinance.com, address: Glarnischstrasse 8, 8002 Zurich, Switzerland.)}
\begingroup\renewcommand\thefootnote{\textsection}
\footnotetext{Equal contribution}
\endgroup
\fi

\begin{abstract}
Blockchain has ushered in a new paradigm, decentralizing economies through the promise of distributed trust within peer-to-peer networks. This very nature of decentralization, pivotal to blockchain's allure, not only bolsters system security but also champions the democratization of systems and procedures. Nonetheless, the blockchain community has yet to converge on a standardized definition or metric for decentralization. Our Systematization of Knowledge (SoK) on blockchain decentralization aims to bridge this definitional void, anchoring its focus on quantification, measurement, and methodological coherence. Our initial endeavors have been directed toward the creation of a taxonomy to elucidate blockchain decentralization. This taxonomy encompasses five integral facets: consensus, network, governance, wealth, and transaction. A deep dive into existing studies reveals an overwhelming tilt towards consensus decentralization, leaving the other facets somewhat in the shadows. To offer a more granular understanding, we present an innovative index grounded in the transformation of Shannon entropy. This index offers insights into the decentralization spectrum across the identified facets. Its efficacy is further fortified through simulations that evaluate its comparative statics. Furthermore, our study doesn't stop at this index; we also venture into alternative decentralization metrics, notably the Gini Coefficient, Nakamoto Coefficient, and Herfindahl-Hirschman Index (HHI). Augmenting its practicality, we have made available an open-source Python utility on GitHub to compute these metrics.

Security research has faced criticism for not adopting commonly accepted scientific methods prevalent in disciplines like medicine, and blockchain decentralization studies are no exception. Our empirical methods — descriptive, predictive, and causal — accentuate the potential of rigorous research approaches in the blockchain realm. Our descriptive analysis has unveiled a trend: over time, decentralization levels seem to gravitate toward convergence. When juxtaposing various DeFi platforms, it emerges that exchange and lending applications manifest higher decentralization compared to payment and derivatives platforms. Predictive insights further reveal a direct correlation between Ether's returns and the degree of transaction decentralization in Ether-backed stablecoins. On the causal front, Ethereum's transition to the EIP-1559 transaction fee mechanism showcases a significant impact on the decentralization dynamics of DeFi transactions.

In summation, our exploration charts a roadmap for prospective blockchain research, emphasizing: 1) the complex interplay and balance among various decentralization facets, 2) the blueprinting of mechanisms that nurture sustained decentralization, and 3) the intricate relationship between degrees of decentralization and the paramount goals of security, privacy, and operational efficiency. We round off our study by spotlighting the challenges that arise in the discourse on blockchain decentralization nuances.
\end{abstract}
%\vspace{-0.5 em} 


\section{Introduction}
Blockchain has revolutionized the financial sector by championing the concept of decentralization \cite{harvey_2021_defi}. At its core, decentralization not only bolsters security but also democratizes systems, offering a genuine distributed peer-to-peer network \cite{puthal_2018_the,budish_2022_the,weyl_2022_decentralized}. While blockchain and decentralized finance (DeFi) applications pledge non-custodial, transparent auditing \cite{schr_2020_decentralized, gomber_2017_digital, werner_2021_sok}, the reality showcases varying degrees of decentralization in different blockchain applications \cite{lin_2021_measuring,oliva_2020_an,gencer_2018_decentralization,decollibus_2021_heterogeneous,cong_2022_inclusion,campajola_2022_the}. Notably, Ao et al. (2022) \cite{ao_2022_are} and Zhang et al. (2022) \cite{zhang2022blockchain} identified a core-periphery dynamic in DeFi token transactions, contrasting the evenly distributed network ideal of blockchain. Heimbach et al. (2023a) \cite{heimbach2023defi} observed a centralizing trend in the transaction graph with the growing popularity of DeFi and NFT marketplaces. Further investigations by Heimbach et al. (2023a) \cite{heimbach2023ethereum} and Wahrstatter (2023) \cite{wahrstatter2023time} unveiled centralization issues in Ethereum's Proof-of-Stake block-building process. While Karakostas et al.\cite{karakostas2022sok} have presented a stratified approach to examine decentralization, the industry still lacks comprehensive and consistent measures for its quantification. Furthermore, Herley \cite{herley2017sok} critiques the security research domain for not embracing standardized scientific methods akin to fields such as medicine, with blockchain decentralization research being no outlier. In this paper, we explore an SoK on blockchain decentralization focusing on measurement, quantification, and scientific methodology by answering the following three questions\footnote{Systemization of Knowledge (SoK) papers evaluate, systematize, and contextualize existing knowledge. Referring to IEEE Symposium on Security and Privacy\url{https://oaklandsok.github.io/} and Journal of Systems Research\url{https://www.jsys.org/type_SoK/}} :  
%“Systemization of Knowledge (SoK) papers evaluate, systematize, and contextualize existing Knowledge. They provide an important new viewpoint on an established, major research area. The heart of the SoK paper is analysis: analyzing the existing literature and providing insights that could not be obtained by simply reading each of the papers. SoK papers analyze the current research landscape: identify areas that have enjoyed much research attention, point out open areas with unsolved challenges, and present a prioritization that can guide researchers to make progress in solving important challenges. a. Qualitative analysis of existing research b. Empirical analysis of open-source software c. Identifying future research directions or challenges that require the community's attention d. Presenting a convincing, comprehensive taxonomy.”

\begin{itemize}
    \item RQ1 (\textbf{establish the taxonomy of blockchain decentralization}). What are the facets of blockchain decentralization and how does existing literature characterize blockchain decentralization in definition and measurement? 
	\item RQ2 (\textbf{propose an explainable decentralization index}). Can we propose an explainable index that measures and quantifies the decentralization level of blockchain across different facets which is generally applicable?
	\item RQ3 (\textbf{apply the decentralization index to empirical research}). Using open-source codes, how can we apply the scientific method of descriptive, predictive, and causal inference to study blockchain decentralization?
\end{itemize}

We commence by examining the literature focusing on the multifaceted nature of blockchain decentralization—both in terms of the blockchain layer's influence and the approach to its measurement. Drawing from this exhaustive review, we endeavor to formulate a taxonomy for blockchain decentralization. Subsequently, our goal is to introduce an interpretable index that quantifies the degree of blockchain decentralization. In the realm of information theory, entropy \cite{shannon_1948_a} serves to quantify the information borne by events and appraises the disparity in a probability distribution.\footnote{Entropy's versatility is evident in its application across diverse fields: from gauging the uncertainty in signal processing \cite{galar_2017_preprocessing}, data mining \cite{mcclean_2003_data, prietoguerrero_2019_nonlinear}, and image detection \cite{wu_2013_local,singh_2008_edge}, to evaluating market volatility in finance \cite{sheraz_2015_entropy,zhou_2013_applications} and determining diversity or inequality in economic systems \cite{mishra_2019_shannon, lim_2020_analysis,teza_2021_entropic,pourmohammadi_2018_evaluating,haynes_2021_recent}. In the blockchain context, entropy has been employed to scrutinize consensus decentralization \cite{gomber_2017_digital,srinivasan_2017_quantifying,kwon_2019_impossibility}.} Adapting entropy to gauge the randomness inherent in transaction distribution, we introduce a novel decentralization index—termed the entropy of transactions. This index possesses the potential for broad application in evaluating other blockchain facets. We assess our index's explicability through comparative statics \cite{dixit1986comparative}, contrasting the simulated decentralization indices from diverse transaction distributions. Furthermore, our study doesn’t stop
at this index; we also venture into alternative decentralization
metrics, notably the Gini Coefficient, Nakamoto Coefficient,
and Herfindahl-Hirschman Index (HHI). Furthermore, by leveraging DeFi data extracted from the Ethereum blockchain, we illustrate the application of three rigorous scientific methodologies—description, prediction, and causal inference—to explore the nuances of blockchain decentralization. Our research culminates in three pivotal findings (F) that directly address the research questions (RQs).
\begin{itemize}
    \item F1. We establish a comprehensive taxonomy for analyzing blockchain decentralization in the five facets of consensus, network, governance, wealth, and transaction regarding the blockchain layer impacted. We find relatively little research on aspects other than consensus decentralization. 
    
    \item F2. We propose a new index that intuitively characterizes the level of decentralization by transforming Shannon entropy. Our index increases when the number of transactions increases and when the distribution of transactions is more evenly distributed, intuitively reflecting greater levels of decentralization. We also define and discuss alternative indices, including the Gini Coefficient, Nakamoto Coefficient, and Herfindahl-Hirschman Index (HHI). We provide open-source Python code for calculating the decentralization index by comparing various alternatives on GitHub.
    
    \item F3. By applying our proposed index to empirical research on top DeFi applications, we demonstrate how the three scientific methods of description, prediction, and causal inference are useful in studying blockchain decentralization. In descriptive visualizations, we find that the level of transaction decentralization varies greatly across different DeFi applications and within an application across time. \footnote{First, decentralized exchanges (DEXs) and lending applications appear slightly more decentralized overall than payment and derivative applications. This is likely due to a greater demand for exchanging and borrowing, leading to a greater number and more even distributions of transactions. Intuitively, applications that become more popular and are utilized more tend to become more decentralized over time, while applications that are utilized less become less decentralized over time. Second, there appears to be a convergence in the level of decentralization across all applications.} In the predictive analysis, we document interactions between the levels of decentralization and economic factors. We discover that a greater return of ether, the native coin of the Ethereum blockchain, predicts a greater decentralization level in stablecoins transfers that include ether as collateral. Lastly, we study how the change in economic mechanism affects the level of decentralization using causal inferences. We discover a significant increase in decentralization in response to Ethereum Improvement Protocol 1559 (EIP-1559), an essential modification to the Ethereum blockchain's transaction fee mechanism.
    
\end{itemize}

We contribute to the four categories that could qualify as a SoK paper. 

\begin{enumerate}
  
    \item [$\blacksquare$] \textbf{Qualitative analysis of existing research}: We have conducted a qualitative analysis of existing research on blockchain decentralization, presenting a comprehensive literature review.
    \item [$\blacksquare$] \textbf{Empirical analysis of open-source software}: We propose a novel decentralization index and apply it to empirical analysis, including description, prediction, and causal inference. We make all our code and data open source. 
    \item [$\blacksquare$] \textbf{Identifying future research directions or challenges that require the community's attention}: We identify future research directions: 1) to explore the interactions between different facets of blockchain decentralization, 2) to design blockchain mechanisms that achieve sustainable decentralization, and 3) to study the interplay of decentralization levels and the ends of security, privacy, and efficiency. We also raise challenges in addressing the controversies in blockchain decentralization.
    \item [$\blacksquare$] \textbf{Presenting a convincing, comprehensive taxonomy}: We establish a taxonomy for analyzing blockchain decentralization in the five facets of consensus, network, governance, wealth, and transaction. 
\end{enumerate}

\section{Taxonomy of Blockchain Decentralization}
We establish a taxonomy of blockchain decentralization after conducting an in-depth literature review by categorizing the facets of decentralization on the corresponding layers of the blockchain system. The current literature on blockchain decentralization examines the impacts of decentralization on various blockchain layers without clearly defining the term "decentralization" and the scope of impact. For instance, the meaning of decentralization is vastly different for the consensus layer, concerned with the security of the blockchain protocol, and for the application layers, concerned with the distribution of usage and wealth. We propose a taxonomy of 5 facets: consensus, network, wealth, governance, and transaction. Table~\ref{tab:16} summarizes the existing blockchain decentralization literature by taxonomy.

The decentralization of a blockchain is manifested within many dimensions due to the layered nature of blockchains. Beyond the core cryptographic infrastructure, the architecture of a blockchain consists of both the underpinning hardware and applications built on top of the blockchain itself. Various blockchain architecture definitions are adopted in the literature to study different aspects of the blockchain ecosystem. A standard definition to analyze blockchain models and security consists of data, network, consensus, incentive, contract, and application layers \cite{zhang_2019_security,yuan_2018_blockchain}. In the analysis of DeFi on blockchains, Werner et al. (2021) \cite{werner_2021_sok} categorizes the blockchain into the base layer and the contract layer, while Schär (2020) \cite{schr_2020_decentralized} proposes settlement, asset, aggregation, protocol, application, and aggregation layers. In light of the existing blockchain decentralization literature and to comprehensively analyze the decentralization of blockchain systems, we further elaborate on the three layers of infrastructure, incentive, and application that consolidate the layers used in the blockchain modeling and security literature. The infrastructure layer consists of data, network, and consensus layers, including data collection into blocks, the communication model (distributed networking, data forwarding, and verification), and the consensus mechanism (e.g., PoW, DPoS). The incentive layer consists of economic rewards to the blockchain system, including incentive mechanisms of the block creation process and token/asset distribution. The application layer consists of contract and application layers, including smart contracts, algorithms, and mechanisms to facilitate high-level applications built on the blockchain system and all blockchain use cases, such as DeFi. 

We contribute to the literature by incorporating blockchain decentralization's definition, measurement, and quantification into the layered architecture. Consensus and network decentralization are connected to the infrastructure layer, wealth decentralization is connected to the incentive layer, and governance and transaction decentralization is connected to the application layer. 
\subsubsection{Consensus Decentralization}
Consensus \cite{zhang_2019_analysis,smith_2021_consensus,cachin_2017_blockchain,bach_2018_comparative} is defined as a state where all the entities or nodes of a blockchain system maintain the same distributed ledger. Consensus is the fundamental mechanism to maintain a blockchain system and information stored within a blockchain \cite{du_2017_a,saleh_2020_blockchain,halaburda_2021_an}.
Eyal (2015) \cite{eyal_2015_the} shows that the decentralization of consensus is crucial to the security of a blockchain system. Cong et al. (2022) \cite{cong_2020_decentralized} and Wu et al. (2019)  \cite{kwu_2019_an} define consensus decentralization as the “evenness” in the distribution of “mining power” in proof of work (PoW) permissionless blockchains.  Liu et al. (2022) \cite{liu_2022_understanding} and Li and Palanisamy (2020) \cite{li_2020_comparison} describe consensus decentralization as the distribution in voting delegation power (the power to delegate creators of blocks) for delegated proof of stake (DPoS) blockchains. 

Consensus decentralization thus characterizes the decentralization in the participation of consensus processes, such as mining PoW protocols and staking or voting in proof of stake (PoS) protocols.  Cong et al. (2022) \cite{cong_2022_inclusion}, Gencer et al. (2018) \cite{gencer_2018_decentralization}, Kwon et al. (2019) \cite{kwon_2019_impossibility},
Li and Palanisamy (2020) \cite{li_2020_comparison}, Lin et al. (2021) \cite{lin_2021_measuring}, Srinivasan (2017) \cite{srinivasan_2017_quantifying} and Wu et al. (2019) \cite{kwu_2019_an}\footnote{Wu et al. (2019) \cite{kwu_2019_an} discover that Bitcoin is 12\% more decentralized than Ethereum in terms of mining power and 9\% more decentralized in terms of wealth.}  empirically measure consensus decentralization using metrics such as the Gini coefficient, Shannon entropy, and the Nakamoto coefficient. Arnosti and Weinberg (2022) \cite{arnosti_2022_bitcoin}, Capponi et al. (2021) \cite{capponi_2021_proofofwork}, Chu and Wang
(2018) \cite{chu_2018_the}, Cong et al. (2020) \cite{cong_2020_decentralized}, Eyal and Sirer (2018) \cite{eyal_2018_majority} and Gervais et al. (2014) \cite{gervais_2014_is} theoretically examine the decentralization of consensus protocols using methods such as game theory to model consensus participation and equilibrium. Xu et al. (2018) \cite{xu2018eos} theoretically examine the decentralization of DPoS-based blockchains and their vulnerabilities. Berman (2018) \cite{berman_2018_eos} and Xu et al. (2018) \cite{xu2018eos} discuss the prevalence of voting collusion on DPoS-based blockchains that lead to centralization.
\subsubsection{Network Decentralization}
Gencer et al. (2018) \cite{gencer_2018_decentralization} describe network decentralization as the fragmentation of control over the network manifested in the security and fairness of a network system. Network decentralization thus depicts decentralization in the underlying blockchain peer-to-peer network infrastructure, such as how users communicate with the network and how nodes communicate with each other. Gencer et al. (2018) \cite{gencer_2018_decentralization} examine network decentralization using prior internet measurement techniques along with the geographical distribution of nodes. Lee et al. (2021) \cite{lee_2021_dq} also study network decentralization through the geographical distribution of nodes. 

\subsubsection{Wealth Decentralization}
Wealth decentralization is commonly defined as the evenness in the distribution of wealth in the blockchain literature \cite{gupta_2018_gini,kwu_2019_an}. Wealth decentralization thus captures the decentralization of monetary assets in tokens and native cryptocurrencies distributed across blockchain users. Srinivasan (2017) \cite{srinivasan_2017_quantifying} defines wealth decentralization as how evenly assets are distributed on the native blockchain or application layer. Gupta and Gupta (2018) \cite{gupta_2018_gini}, Lin et al. (2021) \cite{lin_2021_measuring} apply the Gini coefficient \cite{gini_1909_concentration} to quantify wealth and consensus decentralization levels. \cite{cong_2022_inclusion} utilizes the Herfindahl–Hirschman Index (HHI) to describe the wealth decentralization of Ethereum.
They discover that Bitcoin \cite{gupta_2018_gini} and Ethereum \cite{lin_2021_measuring, cong_2022_inclusion} have very high levels of wealth concentration. 
\subsubsection{Governance Decentralization}
Governance \cite{reijers_2016_governance,pelt_2020_defining} is often defined as the processes of governing, whether undertaken by a government, market, or network, whether over a family, tribe, formal or informal organization, or territory, and whether through laws, norms, power or language. Blockchain governance \cite{pelt_2020_defining,markus_2007_the,gu_2020_empirical,lee_2020_the} is often defined as the means of achieving the direction, control, and coordination of stakeholders within the context of a given blockchain project to which they jointly contribute. Governance decentralization is often defined as the extent to which control over the blockchain is shared between platform owners, and participants \cite{brummer_2019_cryptoassets,chen_2020_decentralized,faguet_2014_decentralization}. Ostrom (1990) \cite{ostrom_1990_governing} describes a fully decentralized governance structure where platform users collectively have complete governance control and can represent their perspectives. Governance decentralization thus specifies the decentralization of ownership and decision-making power on blockchain platforms and how they are shared between owners and participants.  
Although blockchain is, by nature, a system with many decentralized entities, the control and usage of blockchain may not be distributed across many parties. In reality, most blockchain projects have a “foundation” or community of developers with disproportionate influence on decision-making and governing\cite{pelt_2020_defining, wust_2018_do}. Bakos et al. (2019) \cite{bakos_2019_when} further argue that blockchains are often subject to a few parties who operate the system. Pelt et al. (2020) \cite{pelt_2020_defining} define a framework to examine blockchain governance and aspects of governance decentralization.  Chen et al. (2020) \cite{chen_2020_decentralized}\footnote{Chen et al. (2020) \cite{chen_2020_decentralized} assign platforms a “decentralization score” based on data collected from CoinCheckup.com and find that market capitalization often has a U-shaped relationship with governance decentralization, where the best market performance is associated with a middle level of decentralization. The decentralization score is assigned according to the state of governance on CoinCheckup.com, with one denoting Centralized-Hierarchical, two denoting Centralized-Flat, three denoting Semi-Centralized, and four denoting Decentralized.} and Srinivasan (2017) \cite{srinivasan_2017_quantifying} quantitatively measure the decentralization of governance using various proxies, while Gervais et al. (2014) \cite{gervais_2014_is} and Gu et al. (2020) \cite{gu_2020_empirical} qualitatively analyze the governance decentralization of blockchain and blockchain applications. 
\input{tables/tab16}
\subsubsection{Transaction Decentralization}
The measurement of transaction decentralization is mainly absent in the current blockchain literature. In traditional financial markets, trades are executed in a single market clearing, usually controlled by a centralized market maker.  Malamud and Rostek (2017) \cite{malamud_2017_decentralized} describe transactions as "decentralized" when all financial assets in a given market are traded in multiple coexisting and interconnected trading venues. 
 Chu and Wang (2018) \cite{chu_2018_the}, Cong et al. (2022) \cite{cong_2022_inclusion} characterize transaction decentralization as the proportion of transactions conducted by the top nodes in value transacted. Similarly, we define transaction decentralization as the evenness in distributing transactions across blockchain users. This definition further describes the extent of distributed usage among different parties. Transaction decentralization thus describes the decentralization of blockchain-to-peer transactions, especially regarding usage distribution. Cong et al. (2022) \cite{cong_2022_inclusion} further explore transaction decentralization of blockchain networks using utilization and congestion metrics in combination with the network structure analysis.  De Collibus et al. (2021) \cite{decollibus_2021_heterogeneous} examine the concentration of blockchain transaction networks as growth occurs through super-linear preferential attachment, a phenomenon that new nodes of a network have a higher probability of linking with more well-connected nodes, causing a few nodes to become hubs quickly.

\section{Decentralization Index: Definition, Properties, and Simulations}

To fill the gap in the literature on the decentralization of transactions, we propose a new index based on Shannon entropy, which is often used to measure decentralization in areas such as consensus decentralization. The index is adapted from Shannon entropy to measure the degree of randomness in the distribution of transactions, with a higher value indicating more disorder and more decentralization and a lower value of entropy indicating less disorder and more centralization. With a base $2$ exponential transformation, index values range from $1$ to $N$, where $N$ is the total number of transactions.
% Head 2
\subsection{Definition}
We define our decentralization index $H(V)$ as: 
\begin{equation}
\label{eqn:01}
    H(V) = 2^{-\sum_{i=1}^{N}P(v_{i})log_{2}[P(v_{i})]}\\
\end{equation}
or equivalently
\begin{equation}
\label{eqn:02}
        H(V) = \prod_{i=1}^{N}P(v_{i})^{-P(v_{i})}\\
\end{equation}
where $v_i$ is the value of each transaction and $P(v_i)$ is the weight of each transaction concerning the total transaction value.
\begin{equation}
\label{eqn:03}
   P(v_{i})=\frac{v_{i}}{\sum_{i=1}^{N}v_{i}}\\
\end{equation}

\subsection{Properties and Lemmas}
This section explores the properties and lemmas of our decentralization index. They are derived from the properties of Shannon entropy; see Appendix \ref{prop}. 
For simplicity and clarity, we use $p_i$ to denote $P(v_i)$ and denote $H_n(p_1,..., p_n)$ as $H(V)$.
\subsubsection{Properties}
\begin{enumerate}
    \item \emph{\textbf{Property 1 Continuity:}} H should be continuous, such that arbitrarily small changes in the index should be achievable through sufficiently small changes in transaction value. 
    \\
    \item \emph{\textbf{Property 2 Symmetry:}}  $H$ should be unchanged if the outcomes $x_i$ are reordered. That is, $H_n(p_1,p_2,...,p_n) = H_n(p_{i_1},p_{i_2},...,p_{i_n})$ for any permutation ${i_1,...,i_n} \in \{1,...,n\}$.
    \\
    \item \emph{\textbf{Property 3 Maximal at Uniform Distribution:}} $H_n$ should be maximal if all the outcomes are equally likely.  $H_n(p_1,p_2,...,p_n) \le H_n(\frac{1}{n},...,\frac{1}{n})$
    \\
    \item \emph{\textbf{Property 4 Increasing with Number of Transactions: }}For equivalued transactions, the index should increase with the number of transactions. $\underbrace{H_n(\frac{1}{n},...,\frac{1}{n})}_\text{n} < \underbrace{H_{n+1}(\frac{1}{n+1},...,\frac{1}{n+1})}_\text{n+1}$
    \\
    \item \emph{\textbf{Property 5 Multiplicity:}} The index of independent transactions is the product of the index of each event. $H(p_1,p_2,...,p_n)=H(p_1)\times H(p_2)\times ...\times H(p_n)$.
\end{enumerate}
\subsubsection{Lemmas}
Thus, we have the following lemmas based on the stated properties and the properties of Shannon entropy (see the appendix for the properties of Shannon entropy): 
\begin{enumerate}

    \item \bf{\emph{Lemma 1:}} The index is greater than or equal to 1 for any number of transactions, i.e., $H(V)\geq1$.
    \item \emph{Lemma 2:} For a single transaction, the index equals 1, i.e., $H(1)=1$.
    \item \emph{Lemma 3:} The maximal value of the index is $N$, where $N$ is the total number of transactions, i.e., $H(V)\le N$.
\end{enumerate}
\subsection{Simulations}
We designed the following simulations to elucidate better how variation in the number and distribution of transactions affects the decentralization index. Assume there are $N$ total transactions: $1,2,...,n,...,N $ and each transaction has a value $v_n=e^{n \lambda}$, where $\lambda$ is a constant greater than or equal to 0. Therefore, the weight of each transaction is: 
\[P(v_n)=\frac{e^{n\lambda}}{\sum_{n=1}^{N}e^{n\lambda}}\]
\begin{figure}[!htbp]
\centering

       \includegraphics[width=1\linewidth]{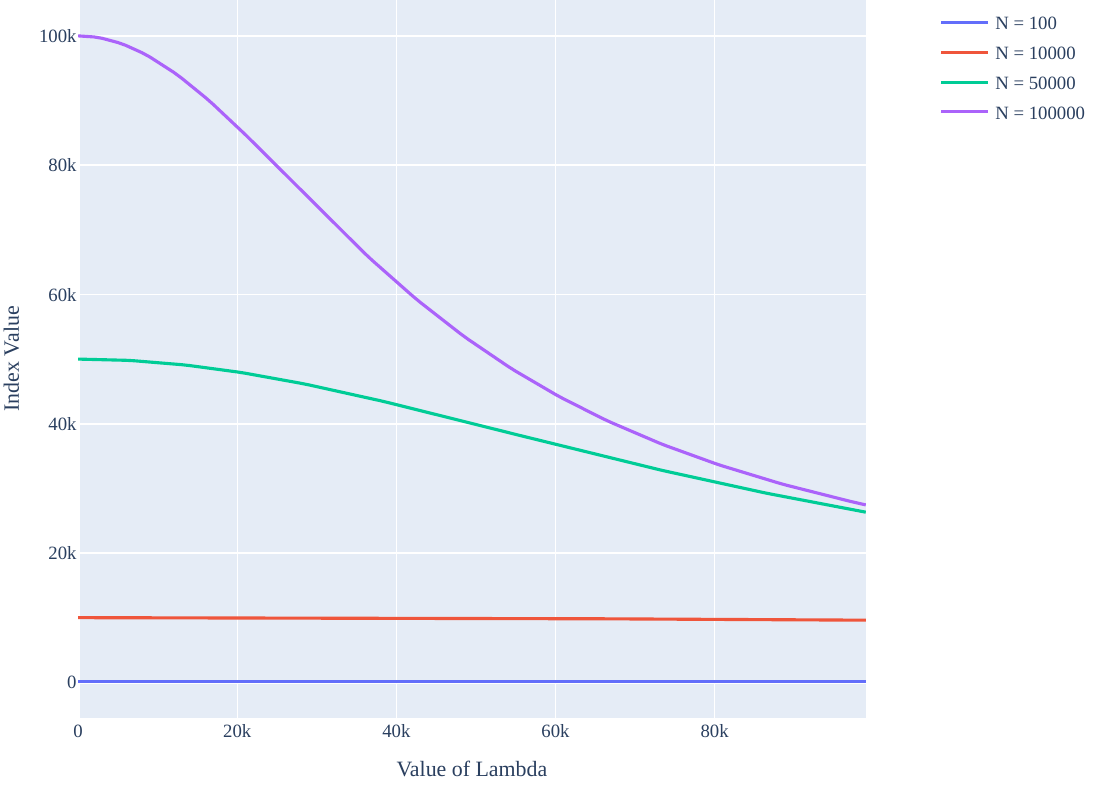}

       \includegraphics[width=1\linewidth]{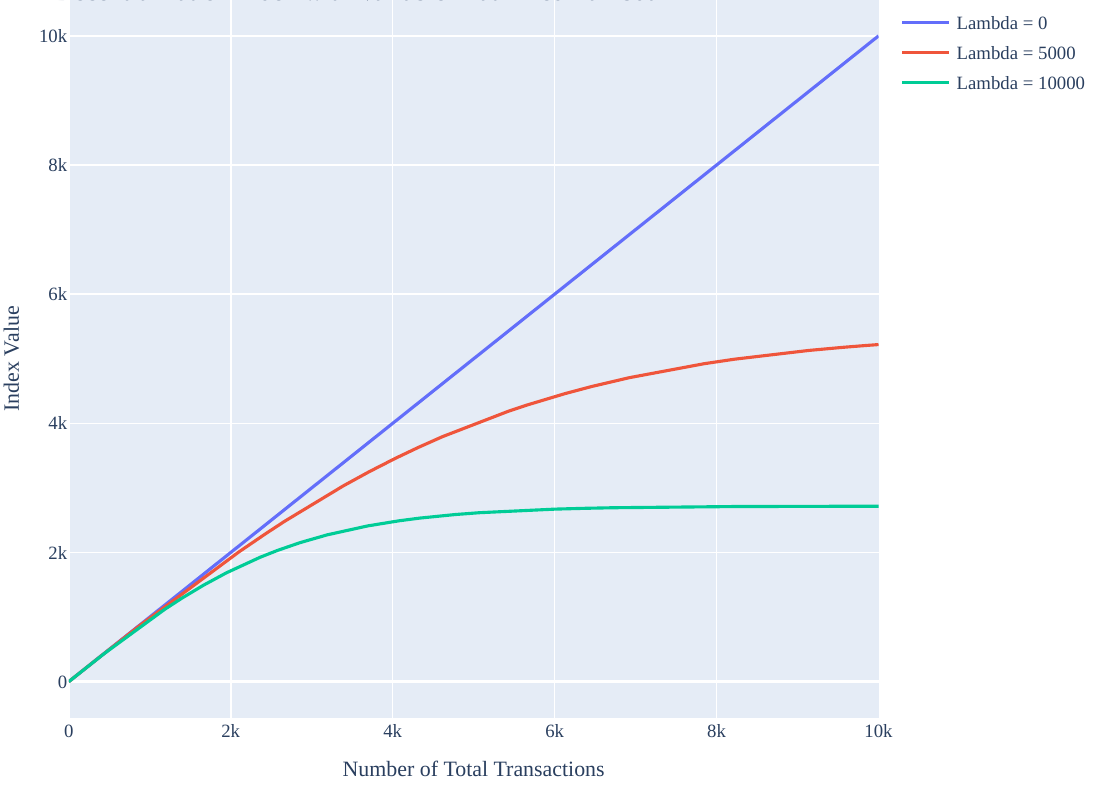}

	\begin{tablenotes}
    \footnotesize
    \item The top graph shows how the decentralization index changes as $N$ \item increases from 0 to 10000 with $\lambda$ fixed at 0, 5000, or 10000. \item The bottom graph shows how the decentralization index changes as \item $\lambda$ increases from 0 to 10000 with $N$ fixed at 100, 10000, 50000, \item or 100000. 
    \end{tablenotes}
	\caption{Simulations of the decentralization index when fixing the total number and distribution of transactions}
	\label{fig:8}
\end{figure}
As explained in  Levin and Zhang (2020) \cite{levin2020}, the distribution function $p(v_n)$ is derived from a specific instance of quantal response equilibrium (QRE), i.e., logit equilibrium (LQRE), which is commonly used in behavioral economics to classify decision making \cite{anderson_2002_the} and is adapted from mixture-of-types models Stahl and Wilson (1995) \cite{stahl_1995_on}. Notably, the function is also used in softmax action selection in reinforcement learning \cite{gao_2018_on,jang_2017_categorical}. The comparative statics in Figure~\ref{fig:8} compare the decentralization index of different numbers or distributions of transactions, ceteris paribus. We observe two intuitive patterns. First, given $\lambda$, as $N$ increases, the value of the index increases. This shows that the index increases as the number of transactions increases, indicating greater decentralization. Second, given $N$, as $\lambda$ increases, the value of the index decreases. Intuitively, when $\lambda$ equals 0, the distribution is uniform, indicating the greatest level of decentralization. When $\lambda$ equals infinity, the weight of the $n$th transaction is 1, indicating total centralization.

\subsection{Alternative Indices}
\subsubsection{Gini Coefficient}
The Gini coefficient\cite{gini_1909_concentration} is frequently employed to measure economic inequality, indicating the distribution of wealth within a given population. In the context of blockchain, the Gini coefficient could be applied to assess the degree of decentralization within a network by gauging the distribution of mining power, wealth, transactions, etc.
\begin{equation*} G=\frac {\sum _{i=1}^{N} \sum _{j=1}^{N} \left |{ p_{i}-p_{j} }\right | }{2 *N \sum _{j=1}^{N}p_{j}} \end{equation*}
where $p_i$ is the fraction of the individual's total wealth or income. In a society where wealth is equally distributed, $G=0$, and in a society where a single person owns 100 percent of the wealth and the remaining $N - 1$ people receive none, $G=1-1/N$.
\subsubsection{Nakamoto Coefficient}
The Nakamoto coefficient refers to the smallest number of entities that must work together to accumulate more than 51\% of the total mining power and compromise a blockchain system. It could be adapted to measure the number of entities required to accumulate 51\%  of wealth, transactions, etc. Unlike the Gini coefficient, which ranges from 0 to 1, the Nakamoto coefficient defines a single threshold, often considered the number of nodes needed to compromise a blockchain system. Chu et al. (2018)\cite{chu_2018_the} further provide a generalized Nakamoto coefficient, $N_{\epsilon}$ that measures the number of entities required to perform $1-\epsilon$ fractions of transactions, which delegates to the Nakamoto coefficient when $\epsilon=0.51$, i.e., the system is operating Nakamoto consensus. 
\begin{equation}
    N = \min \{ k \in [1,\hdots, K]: \sum_{i=1}^{k} p_i \ge 0.51 \}
\end{equation}

\subsubsection{Herfindahl-Hirschman Index (HHI)}
The Herfindahl-Hirschman Index (HHI)\footnote{The index was developed independently by the economists A.O. Hirschman (in 1945) and O.C. Herfindahl (in 1950). Hirschman presented the index in his book, National Power and the Structure of Foreign Trade (Berkeley: University of California Press, 1945). Herfindahl's index was presented in his unpublished doctoral dissertation, "Concentration in the U.S. Steel Industry" (Columbia University, 1950). For more detail on the background of the index, see Albert O. Hirschman, "The Paternity of an Index,"American Economic Review (September 1964), pp. 761-62.} is a measure of market concentration that considers the relative market share of each firm operating within a given industry. The index provides a numerical measure of market concentration that regulators and policymakers use to evaluate potential competition issues within an industry.
\begin{equation}
    HHI = \sum_{i=1}^{N} p_i^2
\end{equation}
where $p_i$ is the market share of the $i$th firm. The index ranges from $1/N$ to 1, where $1/N$ indicates all firms have equal market share and 1 indicates a monopoly.

\subsubsection{Discussion}
There is no consensus on the aforementioned method's applications to measuring decentralization. Each index has its own merits and downsides. For instance, the Gini coefficient has a significant limitation in that it cannot differentiate between different types of inequalities that may exist. In situations where the Lorenz curves intersect, indicating disparate income distribution patterns, they can still result in very similar Gini coefficient values\cite{demaio_2007_income}. The Nakamoto coefficient is a simplified measure that identifies a threshold but does not characterize the underlying distribution. The HHI has similar properties as Shannon entropy. Still, Shannon entropy is more closely related to the number of entities compared to HHI, as it relies more heavily on the many small values of $p_i$. Overall, HHI is influenced more by the larger entities, while the Shannon entropy is more affected by the smaller ones\cite{jost_2006_entropy}. We provide open-source code for all the alternative specifications. As Shannon entropy is commonly used in blockchain decentralization literature to characterize the evenness/unevenness of underlying distributions, we apply the scientific method of descriptive, predictive, and causal inference to study blockchain decentralization using Shannon entropy as an example. Future research can adapt the same methods using alternative indexes.

\section{Data and Empirical Analysis}
\textbf{Data and Code Availability.} We made the data and code open-source, available on GitHub.\footnote{https://github.com/SciEcon/SoK\_Blockchain\_Decentralization}
%\footnote{muted for anonymity}

To show the explainability and applicability of our decentralization index, we conduct descriptive data analysis, event studies, and econometric analysis on major DeFi applications. We also conduct causal inference on EIP-1559 to better understand the impact of economic events on decentralization and user behavior.  We integrate economic theory with an explainable AI approach\cite{adadi_2018_peeking} to ensure our predictions are intuitive and explainable. Although DeFi has become an essential application of blockchain technology, the actual level of decentralization in DeFi is largely unknown \cite{oliva_2020_an}. Our decentralization index shows that the decentralization of different DeFi applications varies significantly in practice.

\subsection{Data Description}
We queried data from Google BigQuery, CoinMetrics, and the DeFi Pulse API. We acquired all on-chain transaction data from the traces table in the Google BigQuery Ethereum dataset (see Appendix \ref{metadata} for metadata). We queried all token transactions from each token's genesis to date (the figures include data from genesis to June 30, 2023). We acquired the ranking and category of DeFi applications ranked in terms of total value locked (TVL) from DeFi Pulse and acquired the market capitalization from Coingecko on July 31, 2021. TVL is the total value of assets stored within a DeFi protocol and is a common "size" measure of DeFi applications. Beyond the many traditional financial metrics, TVL \cite{werner_2021_sok,harvey_2021_defi} is often used to measure the popularity and success of DeFi applications \cite{lehar_2021_decentralized}. To explore the level of decentralization for different DeFi applications and application categories ranked by TVL, we acquired the top 5 decentralized exchanges (DEX), lending, asset applications, and the top 4 derivative and payment applications. Some applications have more than one ERC-20 token used for different purposes.
A plethora of DeFi projects with distinct functions exist within the DeFi ecosystem. Lending, DEX, assets, payment, and derivatives commonly categorize  protocols \cite{harvey_2021_defi,werner_2021_sok,lehar_2021_decentralized}. 
\input{tables/table11}
In Table~\ref{tab:1}, we acquired and categorized the top DeFi tokens in each category. We applied our decentralization to the top applications in each category. 
To better contrast the decentralization of different DeFi applications, we categorized the corresponding token type of each DeFi application into governance, stablecoin, liquidity, dividend, yield farming, and yield aggregator, according to Coingecko \cite{a2021}. Werner et al. (2021) \cite{werner_2021_sok} define governance tokens as tokens developers create to allow token holders to influence governance decisions in a blockchain system, usually representing voting power in a DeFi protocol. They also often have other features along with governance: access to a network, receiving dividends, etc. We cannot analyze the decentralization of governance by applying our index to governance tokens, as it does not capture participation in the governance process. However, governance tokens are likewise tradeable assets with intrinsic value beyond governance (such as protocol access) and often act as stakes/shares in a DeFi application \cite{kiayias_2022_sok}.
A stablecoin is a cryptocurrency designed to have a stable price, typically by pegging its price to a commodity or currency or regulating its supply using algorithms. According to CoinMarketCap, liquidity tokens or liquidity provider tokens (LPs) are tokens issued to liquidity providers on a DEX that run on an automated market maker (AMM) protocol. A dividend token is a token that pays out a portion of platform revenue to holders. According to CoinMarketCap, yield farming is staking or lending crypto assets to generate returns in the form of additional cryptocurrency rewards \cite{vermaak_2020_what}. Yield farming tokens are dealt out by yield farming protocols to incentivize liquidity providers. Yield aggregators are DeFi protocols that automatically help users allocate assets to maximize yield farming returns by searching for the best opportunities and adjusting for factors such as gas prices. A yield aggregator token acts as a claim for the deposit for a yield aggregator protocol. We also categorize each token by its economic usage into general payment tokens, platform tokens, product tokens, and cash-flow-based tokens according to the Palgrave Handbook of FinTech and Blockchain \cite{cong_2021_categories}.
\begin{figure}[!htbp]
	\includegraphics[width = \linewidth]{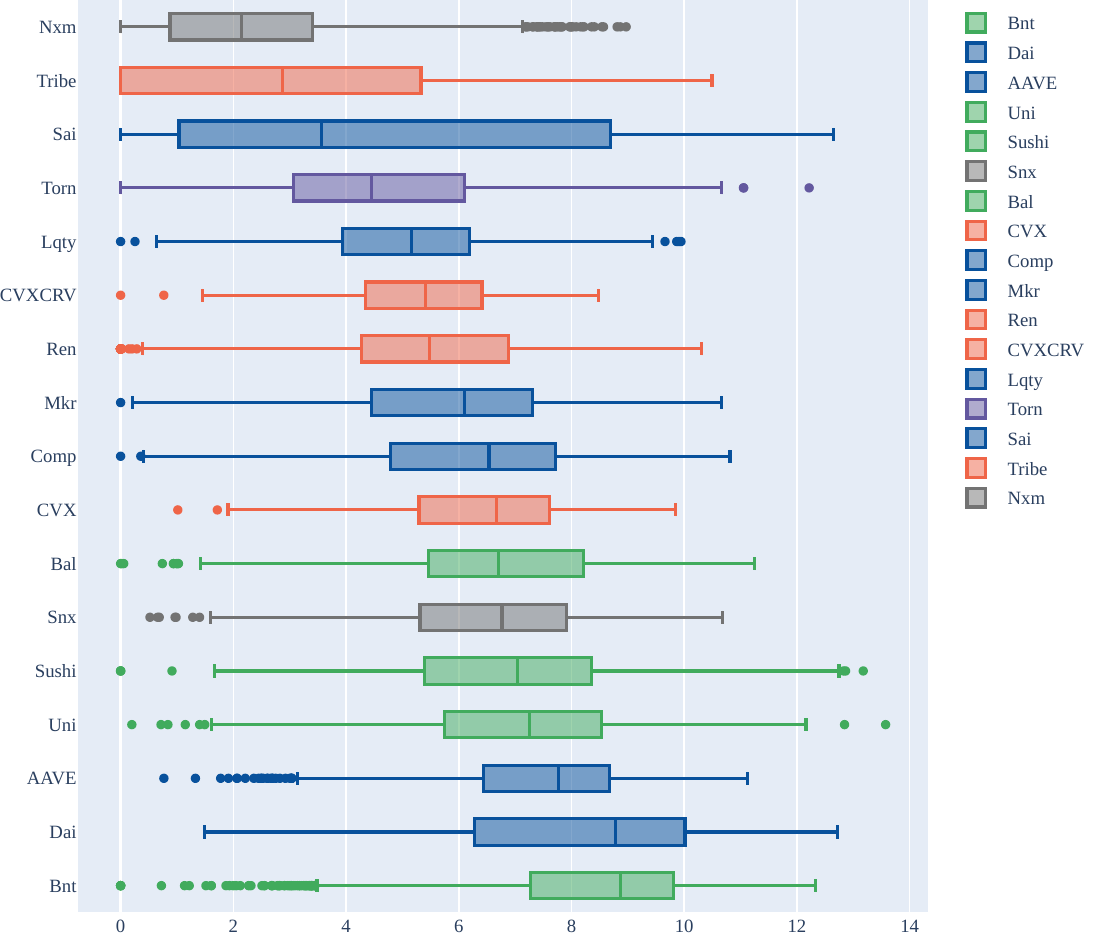}
	\begin{tablenotes}
	\footnotesize
	 \item This figure compares the level of decentralization of top DeFi tokens \item grouped by application category. The decentralization index values of \item each token shown (x-axis) are the base two logarithms of the \item decentralization index (log is taken for visualization purposes). Blue-\item shaded tokens are lending, green-shaded tokens are DEXs, orange- \item shaded tokens are assets, purple-shaded tokens are payment, and \item gray-shaded tokens are derivatives.
	\end{tablenotes}
	\caption{Decentralization Index Distributions of Top Tokens}
	\label{fig:5}
\end{figure}

\subsection{Descriptive Analysis: 
Decentralization Index in DeFi}
To compare the level of decentralization across different tokens and DeFi applications, we create boxplots, histograms, and time-series diagrams. Each instance represents the decentralization index value with a window period of 24 hours. The boxplots show the relative levels of decentralization for different applications and application categories. The histograms elaborate on the distribution of daily decentralization levels within categories. The time-series diagram shows the changes in decentralization values over time, capturing trends and fluctuations due to market dynamics, and aims to illustrate how decentralization is affected by real-world events. 

\begin{figure}
	\includegraphics[width = \linewidth]{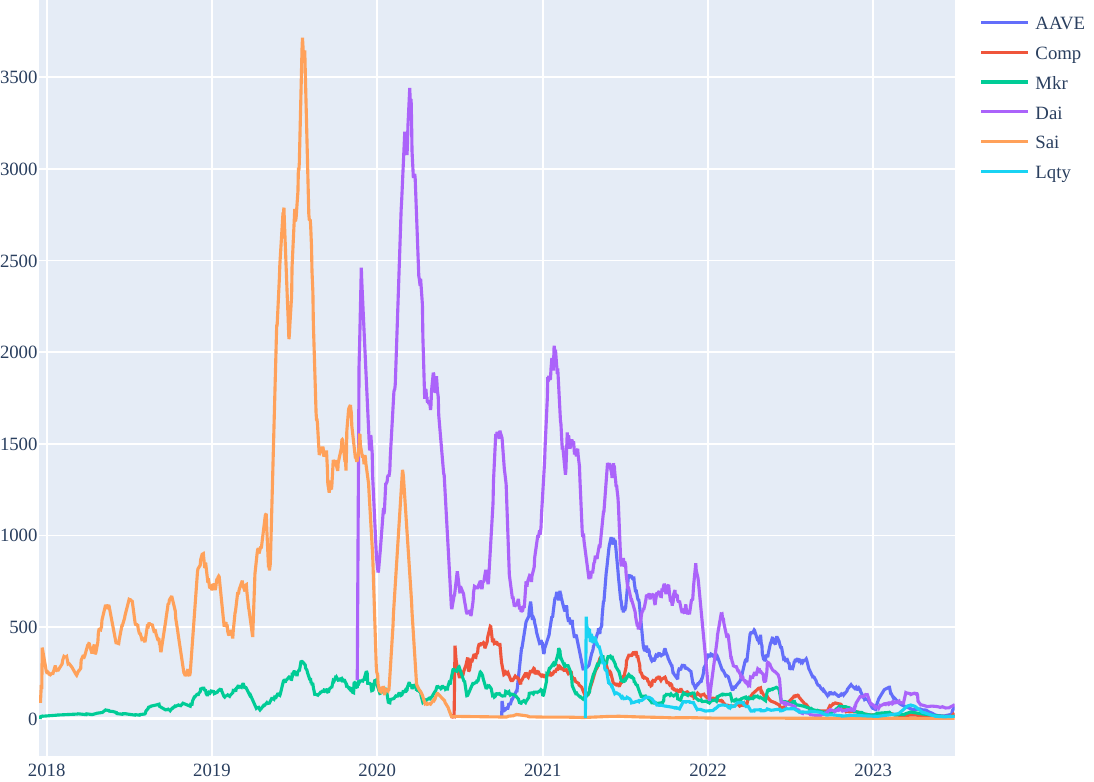}
	\begin{tablenotes}
	\footnotesize
	 \item This figure compares the level of decentralization of top lending \item tokens across time. The figure plots the 30-day simple moving \item average  value of the index. The index values have a daily granularity. \item The  y-axis  is the index value, and the x-axis is the date.
	\end{tablenotes}
	\caption{Decentralization Index of Top Lending Applications}
	\label{fig:2}
\end{figure}

Figure ~\ref{fig:5} shows that DEX and lending applications generally have greater decentralization than payment and derivative applications. This is, in part, due to the number of transactions. DEX and lending applications have, by and large, more daily transactions because the market has a higher demand for exchanging and borrowing than other motives.  
\begin{figure}
	\includegraphics[width = \linewidth]{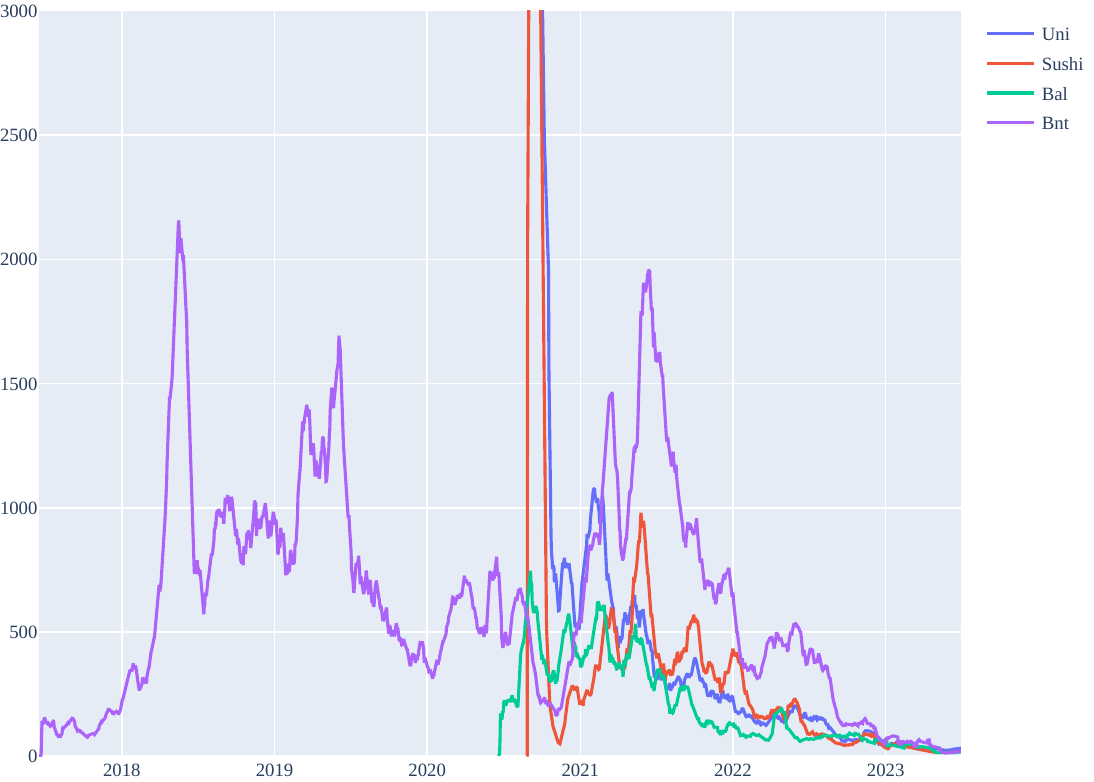}
	\begin{tablenotes}
	\footnotesize
	 \item This figure compares the level of decentralization of top DEX tokens \item across time. The figure plots the 30-day simple moving average value \item of the index. The index values have a daily granularity. The y-axis \item is the index value, and the x-axis is the date.
	\end{tablenotes}
	\caption{Decentralization Index of Top DEX Applications}
	\label{fig:3}
\end{figure}
In Figure ~\ref{fig:3}, we note that Uniswap and Sushiswap have different starting points than Balancer and Bancor because  the former two projects airdropped their native tokens to loads of early users. Despite the very decentralized start, these projects tend to revert to being more centralized and attain a similar level of decentralization over time. Notably, Campajola et al. (2022) \cite{campajola_2022_the} document a similar shift towards greater centralization of transactions measured by the average number of addresses per entity, with a few entities participating in many transactions. Ao et al. (2022) \cite{ao_2022_are} also note an intertemporal trend of centralization using social network analysis on the DeFi token of AAVE, where transactions have become more centralized over time despite an initial decentralized structure. One potential explanation is that most of those who obtained the airdropped tokens will eventually sell their tokens, and the whale holders will accumulate tokens. Regardless of the initial allocation, the token distribution will converge to a similar level. This phenomenon also holds for lending applications (see Figure ~\ref{fig:2}). Cong et al. (2022) \cite{cong_2022_inclusion} corroborate our results, observing a shift from peer-to-peer interactions to more transactions by large players. The mechanism is interesting to explore; we leave it for future research.

\subsection{Predictive Analysis: Market Movements}

\subsubsection{Dai/Sai Stablecoin: Event Analysis}
We use our decentralization index to qualitatively and quantitatively analyze Dai/Sai stablecoin via event and econometric analysis. 
Dai is the largest crypto-backed stablecoin by market capitalization on the Ethereum network and is a key component of the Maker protocol. Dai is a crypto asset-backed stablecoin that accepts multiple types of collateral. It was officially launched on the leading Ethereum network on 18 December 2017. The first version accepts only ether as collateral. On 18 November 2019, the multi-collateral version of Dai was launched, and the original single-collateral version was renamed Sai. Sai was slowly phased out after the launch of Dai, so the special transitory nature of this coin makes it an ideal candidate to investigate the changes in levels of decentralization as user dynamics vary, that is, how decentralization changes as users of Sai transition to Dai. 
Furthermore, stablecoins are often used as a medium of exchange in the highly volatile cryptocurrency market. As an integral component of the DeFi ecosystem, stablecoins are actively involved in day-to-day trading. We seek to quantitatively analyze the impact of cryptocurrency market dynamics on transaction decentralization through the lens of Sai/Dai.
\begin{figure}
	\includegraphics[width = \linewidth]{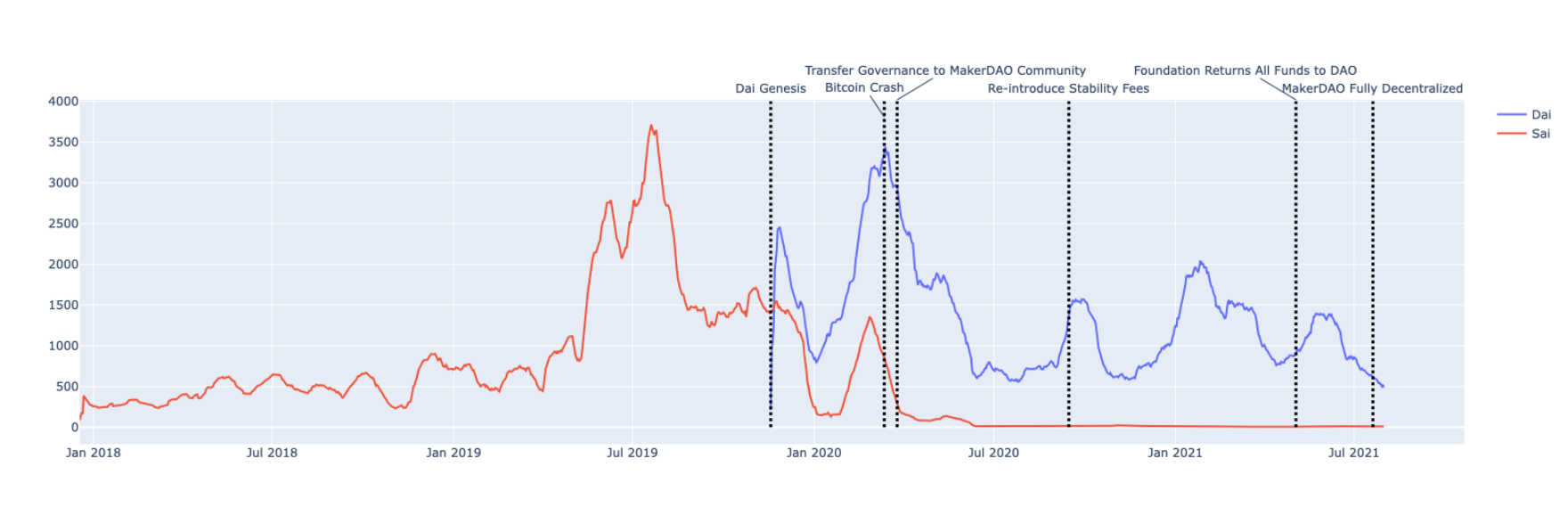}
	\begin{tablenotes}
	\footnotesize
	 \item This figure compares the level of decentralization of Sai (red) and \item Dai (blue) tokens across time and labels important market and \item governance events. The figure plots the index's 30-day simple moving \item average value. The index values have a daily \item granularity. The y-axis is the index value, and the x-axis is the date.
	\end{tablenotes}
	\caption{Sai and Dai Decentralization Index and Historical Events}
	\label{fig:6}
\end{figure}

The impact of user dynamics on decentralization is clearly illustrated in the decline of Sai and the rise of Dai. Figure~\ref{fig:6} shows that the decentralization index of Sai has been decreasing since the genesis of the Dai token, which is the upgraded version of Sai. Sai token holders could irreversibly convert their Sai tokens to Dai tokens, resulting in fewer and more Sai transactions. Such a trend led to a lower decentralization index for Sai but a higher one for Dai.

Furthermore, we look at the impact of specific governance and market events on decentralization (see Table ~\ref{tab:9} in Appendix for event details). We note that events related to governance do not change the decentralization index substantially, as token holders are less sensitive to governance incentives than to economic incentives, such as the reintroduction of stability fees.
We make two important observations about events.
\begin{itemize}
  
  \item The decentralization index is largely affected by the crypto market movement. For example, in March 2020, the cryptocurrency market underwent a large correction. Many token holders flooded into exchanges to trade their tokens. This caused a spike in the decentralization index of both Dai and Sai. This motivates our econometric analysis to quantify the impact of market movement on decentralization.

  \item As the market becomes more volatile, the abrupt price adjustment could trigger the liquidation of collateralized positions in DeFi applications, such as Maker and Liquity. To prevent such an occurrence, there are generally more transactions, such as lowering debt position and topping up the margin, during periods of market turmoil. 

\end{itemize}

On the one hand, the decentralization index increases due to the increasing number of transactions. On the other hand, there are more activities from whale token holders, which lowers the decentralization index. In the next section, we perform an econometric analysis to test which effect dominates.

\subsubsection{Econometric Analysis}
As noted in our event analysis, the market movement appears to affect decentralization substantially. The cryptocurrency market includes both native coins of blockchains, such as Bitcoin and Ether, and tokens that exist on blockchains, such as Dai. Bitcoin and Ether are two assets with the highest market capitalization and are strong representatives of the cryptocurrency market. Sai/Dai is built on the Ethereum blockchain and mostly accepts Ethereum-based assets as collateral, with Sai accepting only ether. Thus, we focus our analysis on Ethereum, Bitcoin, and Sai/Dai. We conducted an econometric analysis on Ether and Bitcoin returns and volatility associated with Sai/Dai stablecoin's decentralization to examine the relationship between the cryptocurrency market movement and our decentralization index. The dependent variable is our decentralization index value, and the independent variables are Ether and Bitcoin returns for different periods, the first principal component (PC) of all returns, and 30-day volatility. All the variables in our analysis have a daily frequency. To ensure the stationary of our time series predictor and response variables used for regression analysis, we conducted an augmented Dickey-Fuller test for stationarity at the  95\% confidence level \cite{dickey_1979_distribution}. The null hypothesis states a unit root in our time-series variable, and the alternative hypothesis states no unit root, indicating stationarity \cite{elliott_1996_efficient}. All the variables used in our analysis are stationary at the $\alpha=5\%$ significance level. We have also applied the Newy-West Estimator for standard errors given the high autocorrelation in our dependent variable %(see Figure ~\ref{fig:9} in Appendix \ref{addfig}).Table ~\ref{tab:19} in Appendix \ref{additionalsts} shows that our decentralization index has high single-period autocorrelation. Bitcoin (BTC) and Ether (ETH) returns are highly correlated across all periods, so we only show the regression results of Ether in this section (see Figure~\ref{fig:10} in Appendix \ref{addfig}, see Appendix \ref{additionalsts} for additional statistical results). Given the colinearity of return variables $\rho > 0.5$ of different timespans and the first principal component (PC) of all return variables, we regressed each variable on the decentralization index of Sai and Dai.
 \input{tables/table20}

Column (1) of Table~\ref{tab:20} shows the relationship between returns/volatility and transaction decentralization of Sai. Column (2) of Table~\ref{tab:20} shows the relationship between returns/volatility and transaction decentralization of Dai. ETH returns have a strong positive relationship with the decentralization of Sai, especially for longer periods. This result suggests that with higher returns, there are higher levels of utilization and trading in Sai and more decentralized market participation. Campajola et al. (2022) \cite{campajola_2022_the} also found that user behavior measured by the average number of addresses per entity is closely related to market returns. Moreover, ETH 30-day volatility of returns has a strong negative relationship with the decentralization of Sai. Intuitively, with higher return volatility, more whale token holders move their assets, decreasing the value of the decentralization index. However, these significant results for Sai do not hold for Dai in the short term, likely due to the nature of Dai, which accepts a wide range of assets as collateral, in contrast to Sai, which accepts only Ether. Despite correlations between crypto assets and a diverse pool of assets, ETH returns do not impact the decentralization of Dai to the same extent.
Our econometric results confirm that the level of decentralization is closely associated with market returns, especially in the mid to long-term. Higher returns are typically related to greater transaction decentralization. This suggests greater market participation overall. The results further answer the question proposed in the Sai/Dai event analysis section: what are the relative magnitudes of influence on the decentralization index between the increase in transactions and the existence of whale holders in turbulent markets? The impact of whale holders and large transactions outweighs the rise in the number of transactions.

\subsubsection{Implications}
Our study utilizes Shannon entropy from information theory to devise a decentralization index specifically for token transactions on the Ethereum blockchain. This index is both interpretable and measurable, adeptly characterizing the intricate dynamics of decentralization. Moreover, our approach highlights variations in transaction decentralization as they relate to market forces. We've discerned that while market returns and volatility significantly influence transaction decentralization, nonmarket forces like governance events do not. Beyond its current application, future research could employ our method to craft (alternative) indices for assessing the decentralization levels of other facets within the blockchain realm. This novel index thereby offers fresh perspectives for probing token economics on blockchains.

%Moreover, our index differs from traditional market transaction measures, such as volume and count. Figure~\ref{fig:12} shows the correlation between transaction volume, transaction count, and our index and the Pearson test results. Although associated with transaction count and volume, our index further characterizes the distribution of transactions not captured by volume or number of transactions. 

\subsection{Causal Inference: EIP-1559}

EIP-1559 is a major overhaul of the transaction fee mechanism (TFM) for the Ethereum blockchain \cite{buterin_2019_eip1559}. EIP-1559 makes Ethereum the first major blockchain to move away from the classic first-price auction mechanism in coordinating the demand and supply of transactions. On August 5th, 2021, Ethereum activated the London Hard-fork, bringing major changes to the TFM by adding a base fee parameter and how users specify the bids for transaction fees. The base fee adjusts dynamically to reflect the minimum gas price users must pay in each block, responding to the block gas used in the previous block. Users can now bid the maximum priority charge per gas and the maximum fee per gas in their transactions. Users tip miners to prioritize their transactions using priority fees per gas unit. The maximum fee per gas parameter acts as a cap on both the base and priority fees.  Liu et al. (2022) \cite{liu_2022_empirical} evidence the positive impacts of EIP-1559 on user experience by simplifying the fee estimation process, reducing the difference in gas price paid between blocks, and decreasing waiting times. These changes smoothen the transaction process, making peer-to-peer transactions easier and swifter. \cite{zhang2023understand} show that the Merge further reduces waiting time and market congestion mainly due to the shortening of block intervals. In this section, we seek the causal impacts of EIP-1559 on transaction decentralization. Our study further connects economic factors to transaction decentralization in cases of mechanism changes. 

\begin{figure}
	\includegraphics[width = \linewidth]{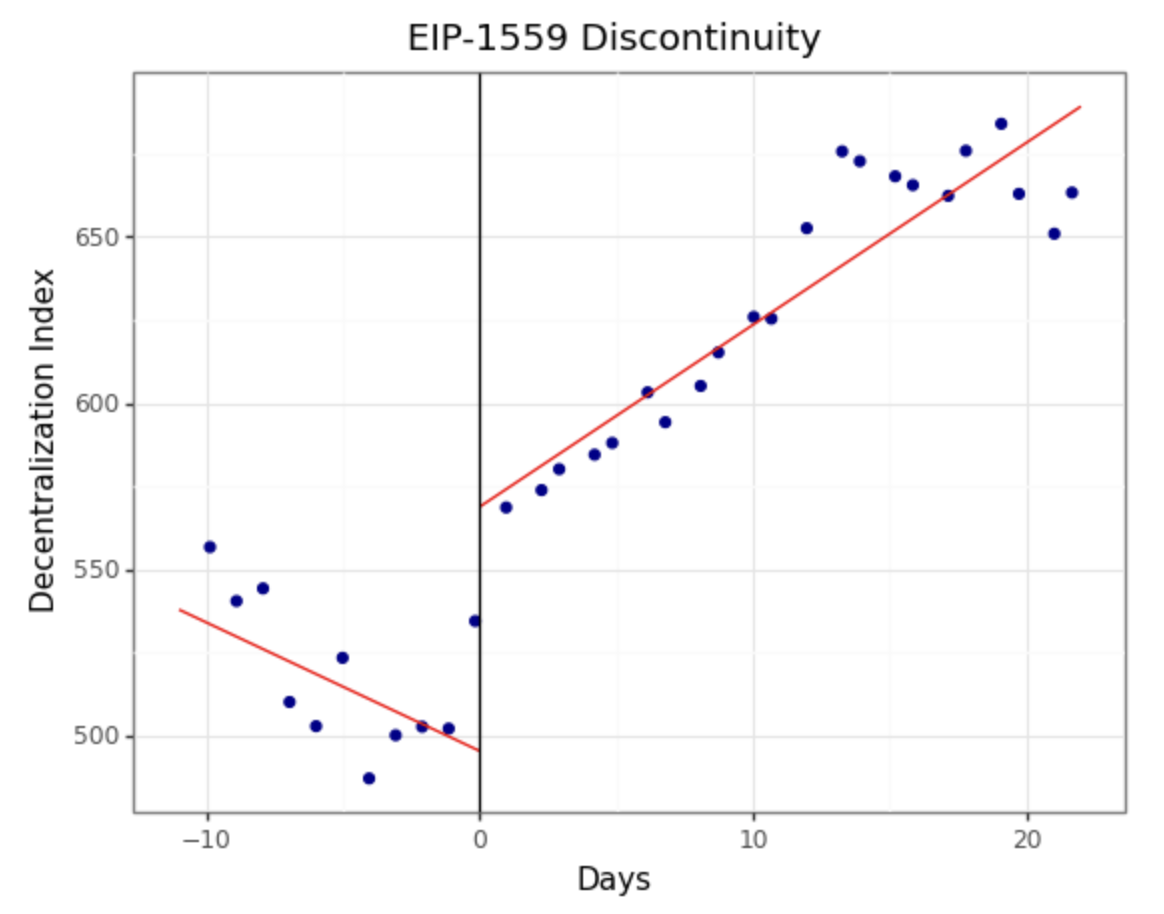}
	\begin{tablenotes}
    \footnotesize
    \item This figure shows the regression discontinuity of Dai during the time \item  frame  of our  EIP-1559 analysis. The figure plots the 30-day simple \item  moving average value of the decentralization index concerning \item  time. The index values have a daily granularity. The y-axis is the index \item  value, and the x-axis is the days from EIP-1559.
    \end{tablenotes}
	\caption{EIP-1559: Discontinuity}
	\label{fig:15}
\end{figure}

\subsubsection{Methodology}
One of the most difficult challenges in determining the causal effect of EIP-1559 on blockchain decentralization is distinguishing the effect of EIP-1559 from confounding factors such as price, volatility, market movement, and time trend. We adopt Regression Discontinuity Design (RDD) \cite{imbens_2008_regression,lee_2010_regression,athey_2017_the} to analyze the effects of EIP-1559 on transaction decentralization. RDD is a quasi-experimental evaluation frequently used in economics, political science, epidemiology, and related fields. We use the London Hardfork, the days from the EIP-1559 London Hardfork, and their interaction as the independent variables in the RDD framework to estimate the immediate effects of the London Hardfork and the average treatment effects of EIP-1559 adoption. Liu et al. (2022) \cite{liu_2022_empirical} adopted a similar framework to analyze the impact of EIP-1559 on various transaction-related economic factors, including waiting times and gas prices, while \cite{cong_2022_inclusion} also uses RDD to analyze the impact of EIP-1559 on mining reward distribution and transaction volume. 

In Eq.~\ref{eqn:06}, $\alpha_1$ is the coefficient that represents the immediate effect (local average treatment effect) of the London Hardfork, $\alpha_3$ is the coefficient that represents the gradual effect of the London Hardfork. 
\begin{equation}
\label{eqn:06}
\begin{split}
   Y = \alpha_0 &+ \alpha_1 \mathds{1}(\text{LONDON HARDFORK}) + \alpha_2 X_{\text{Days}} \\
   &+ \alpha_3 \mathds{1}(\text{LONDON HARDFORK}) \times X_{\text{Days}} + \alpha_4 \pmb{Z} + \epsilon
\end{split}
\end{equation}
We include a set of control variables represented as Z in Eq.~\ref{eqn:06}, including transaction volume, transaction count, return on investment, and 30-day volatility. The study period is from July 25, 2021, to August 27, 2021, including 70 thousand blocks before and 140 thousand blocks after the London Hardfork on August 5, 2021. 
We utilized the RDRobust package\cite{calonico_2017_rdrobust} to generate figures for this section.

\subsubsection{Results and Discussion}
\input{tables/table18}
EIP-1559 significantly impacts decentralization, increasing immediately after the London Hardfork and gradually as a trend in the days after the London Hardfork for stablecoins Dai and USDt. However, EIP-1559 significantly negatively impacts decentralization both immediately and as a trend for lending applications Compound and Aave. Nonetheless, this shows transaction decentralization's intense sensitivity to changes in economic incentive mechanisms and confirms our conclusion on market movements' strong effect on decentralization. Furthermore, EIP-1559's contrasting impact on decentralization for different application categories is interesting. We leave this phenomenon for future research.
Table~\ref{tab:18} shows the regression results for Dai with the 30-day simple moving average of the decentralization index as the dependent variable. Column (1) includes the indicator of London Hardfork, days from the London Hardfork, and their interaction as independent variables. Column (2) adds control variables transaction volume and transaction count. Column (3) further adds the 1-day return on investment and 30-day volatility of returns as controls. 
%We also included the exponential moving average (EMA) with $\alpha = 0.1$ of our decentralization index, an alternate dependent variable, yielding consistent results (Table ~\ref{tab:12} Appendix \ref{additionalsts}). 
We analyzed four assets, including two stablecoins: Dai and USDt, and two lending protocols: Aave and Comp (see Appendix \ref{additionalsts}). The stablecoins produced consistent results showing EIP-1559's positive impact on decentralization; In contrast, the lending protocols produced consistent results showing EIP-1559's negative impact on decentralization.

Figure~\ref{fig:15} shows an immediate and gradual increase in decentralization after the London Hardfork as discussed in the regression results. Notably, this temporary alteration in decentralization caused by EIP-1559 does not reverse the trend of overall convergence towards centralization intertemporally. 

\subsubsection{Sensitivity and Robustness Checks}

We conducted sensitivity checks by estimating the effect twice (Figure~\ref{fig:16}) and half (Figure~\ref{fig:17}) of our defined study period (bandwidth), respectively. The results with twice the original bandwidth (in appendix ~\ref{robustness}) show consistent trends in the immediate increase in decentralization after the London Hardfork and a gradual increase in the following days. Although the results with half the original bandwidth show an immediate increase in decentralization after the London Hardfork is consistent with our results, the increasing trend follows the event initiated before the London Hardfork. We hypothesize that the rational expectations of EIP-1559 influenced decentralization levels in the days immediately preceding the London Hardfork. Nonetheless, our sensitivity checks confirm increased decentralization immediately due to EIP-1559. We also conducted robustness checks by controlling for the probability of adopting EIP-1559 (probability of treatment) via adoption rates. We exploit the gradual and independent adoption of EIP-1559 after the London Hardfork to estimate the immediate and average treatment effect of EIP-1559. The independent variable: days from the London Hardfork, was removed due to high colinearity with adoption rates %(see Figure~\ref{fig:14} in the appendix). 
The results (in appendix~\ref{robustness}) are consistent with our initial setup, registering an immediate increase in decentralization after the London Hardfork event and a gradual increase with increasing adoption of EIP-1559.

\section{Discussion on Future Research and Related Literature}
Blockchain creates a system of decentralized trust that relies on economic incentives to secure the system by ensuring the benefits of dishonesty are outweighed by the costs \cite{NBERw24717}. Decentralization is key in balancing the economic incentives that secure a blockchain from majority attacks. For instance, Weyl et al. (2022) \cite{weyl_2022_decentralized} envisions a decentralized society where commitments, credentials, and affiliations are encoded through non-transferable “soul-bound” tokens (SBTs) maintained on blockchain that creates a trusted network of the real economy. Decentralization lies at the core of what blockchain promises. We identify three future research directions: 
\begin{itemize}
    \item Delve into the nuanced relationships between various aspects of blockchain decentralization and extend the insights to facets beyond the existing literature, e.g. data layer, and identity layer as in Ernstberger et al.\cite{ernstberger2023sok}. For example, Bakos et al.\cite{bakos2021permissioned} posit that while permissioned systems may exhibit greater centralization in terms of access control, they could potentially offer more decentralized governance than their permissionless counterparts. 
    \item Craft blockchain mechanisms aimed at achieving enduring decentralization. Merely incorporating decentralization in protocol design doesn't ensure its manifestation during actual usage. Human behavior is influenced by rationality and psychological factors, as noted by Glazer et al.\cite{glazer2016models}. Consequently, future system architectures must take into account both economic incentives and behavioral heuristics to ensure long-lasting decentralization. 
    \item Examine the intricate relationship between levels of decentralization and the goals of security, privacy, and efficiency. For example, the decentralization of transactions may be essential for the security of a blockchain ecosystem that hosts smart contracts. Consider the Ampleforth stablecoin, which modifies its supply in response to price deviations from its target value\cite{zhang_2021_optimal}. Future research could explore how enhancing transaction decentralization can mitigate malicious manipulation of Ampleforth prices, thereby bolstering the stability of the stablecoin system.
\end{itemize}

\begin{figure}
	\includegraphics[width = \linewidth ]{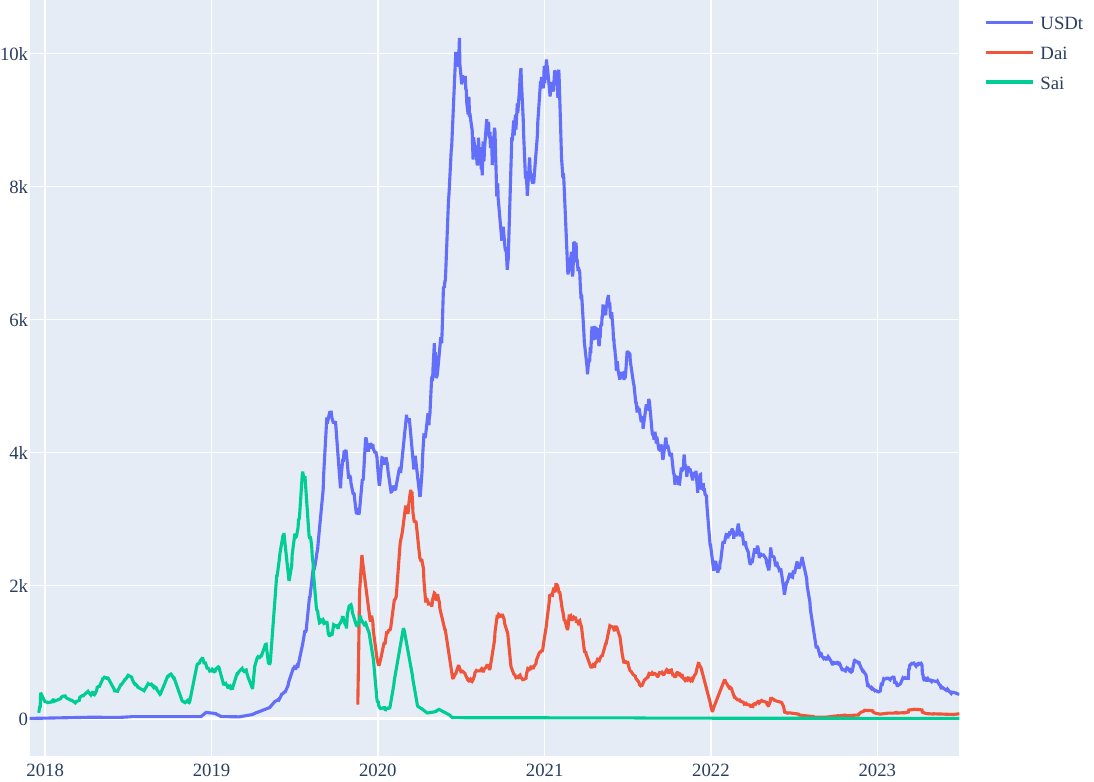}
	\begin{tablenotes}
	\footnotesize
	 \item This figure compares the level of decentralization of USDt (green) \item  Sai (red) and  Dai (blue) tokens across time. The figure plots the \item  30-day simple moving average value of the index. The index values \item  have a daily granularity. The y-axis is the index value, and the \item  x-axis is the date.
	\end{tablenotes}
	\caption{USDt Decentralization Index Comparison with Sai and Dai}
	\label{fig:6.1}
\end{figure}

We highlight the challenges associated with addressing the complexities of blockchain decentralization. For example, Figure~\ref{fig:6.1} depicts a time series comparison of the decentralization indices for the fiat-backed stablecoin USDt and the crypto-backed stablecoins Sai and Dai. While USDt exhibits greater centralization in consensus, intriguingly, it demonstrates more decentralization at the transaction level. What factors contribute to these contrasting decentralization levels between consensus and transaction layers? Moreover, how can we refine measurements of decentralization in the realms of blockchain interoperability and cross-chain solutions\cite{Rafael_survey_2022,Rafael_solution_2022}? We earmark these queries for subsequent research endeavors.

\subsection{Contribution to Literature: Beyond Security to Interdisciplinary Engagement}

%\begin{figure}
	%\includegraphics[width = \linewidth]{figures/fig1.pdf}
	%\begin{tablenotes}
	%\footnotesize
	 %\item This figure summarizes the related literature in the realm of \item blockchain  decentralization, market analysis, and the blockchain \item  ecosystem, and our contribution is shown in \textcolor{red}{red}. 
	%\end{tablenotes}
	%\caption{Related Literature}
	%\label{fig:1}
%\end{figure}

Our primary aim is to extend the reach of blockchain research to interdisciplinary audiences, transcending the confines of the security field. In this light, our research intersects with three key literature domains: the evaluation of blockchain ecosystems with a keen focus on the blockchain trilemma\cite{schr_2020_decentralized,gomber_2017_digital,werner_2021_sok,bakos_2019_when} and the quantification of blockchain decentralization\cite{cong_2020_decentralized,gupta_2018_gini,srinivasan_2017_quantifying,abdulhakeem_2021_powered, kwu_2019_an,lin_2021_measuring}; comprehensive blockchain-related SoK or surveys; and the intricate market dynamics of DeFi\cite{chen_2020_decentralized,harvey_2021_defi,angeris_2021_an,lehar_2021_decentralized}.

%Our research contributes to three strands of the literature: research evaluating blockchain ecosystems, especially  the blockchain trilemma\cite{schr_2020_decentralized,gomber_2017_digital,werner_2021_sok,bakos_2019_when} and the measurement of blockchain decentralization \cite{cong_2020_decentralized,gupta_2018_gini,srinivasan_2017_quantifying,abdulhakeem_2021_powered, kwu_2019_an,lin_2021_measuring}, blockchain-related SoK or surveys, and the market analysis of DeFi\cite{chen_2020_decentralized,harvey_2021_defi,angeris_2021_an,lehar_2021_decentralized}.
\subsubsection{The Blockchain Trilemma}
Decentralization, along with scalability \cite{li_2017_a,prashanthjoshi_2018_a} and security \cite{herrerajoancomart_2016_privacy,zhou_2020_solutions}, are desirable properties of a blockchain ecosystem. However, Vitalik Butterin \cite{hafid_2020_scaling} first described a blockchain trilemma where trade-offs among the three desirable properties are inevitable. The impossible trinity is now widely known in the blockchain community \cite{conti_2019_blockchain,hafid_2020_scaling,team_2019_trifecta} and in crypto-economics \cite{schilling_2020_central, halaburda_2020_the,auer_2021_permissioned}. Thus, when making blockchain design choices, it is critical to measure the level of each desirable property and understand its effect on the ecosystem. We contribute to the literature from the decentralization aspect and analyze the specific SoK topics.

\subsubsection{Blockchain-Related SoK and Surveys}

Our research supplements the vast literature on Blockchain-Related SoK and Surveys. Key concepts covered include Privacy\cite{almashaqbeh_2021_sok,bonneau_2015_sok}, emphasizing confidentiality and payment anonymity. Consensus\cite{bonneau_2015_sok,wang_2021_sok_b}, ensuring ledger maintenance and blockchain security, and Accountability\cite{chatzigiannis_2021_sok}, which focuses on blockchain network auditability. The decentralized nature of blockchains has led to challenges like Front-running\cite{eskandari_2019_sok}, though solutions are emerging\cite{baum_2021_sok}. Foundational to blockchain security is Cryptography\cite{raikwar_2019_sok}. Alongside blockchain attacks, DeFi applications have also seen threats\cite{zhou_2022_sok}, with recent defenses against transaction reordering attacks\cite{heimbach_2022_sok}. Governance\cite{kiayias_2022_sok} remains essential for decentralization and blockchain evolution. Scalability\cite{mccorry_2021_sok} is a persistent challenge, with solutions like Sharding\cite{wang_2019_sok} emerging. Communication\cite{zamyatin_2021_sok} facilitates scaling and inter-blockchain interactions, with works like Wang (2021) \cite{wang_2021_sok} shedding light on interoperability. Rapidly growing, DeFi \cite{bartoletti_2020_sok,werner_2021_sok,xu_2021_sok} exemplifies blockchain's transformative impact on finance. Layer-2 protocols\cite{gudgeon_2020_sok} enhance blockchain operations, while Shi et al. (2021) \cite{shi_2021_when} and Abuidris et al. (2019) \cite{abuidris_2019_a} delve into blockchain applications such as auctions and voting. The blockchain's role in the metaverse\cite{gadekallu_2022_blockchain} and its other potential applications\cite{hassan_2019_blockchain, ruoti_2019_sok} are also discussed. For a broader perspective on blockchain-related SoKs, we direct readers to Zhang (2023) \cite{zhang2023design}. Our work uniquely addresses blockchain decentralization, diving deep into its measurement, quantification, and methodology amidst a plethora of blockchain studies.

\subsubsection{Market Analysis of DeFi}

Crypto assets, due to their distinct nature, have given rise to an extensive body of literature examining cryptocurrency pricing, valuation, and market return patterns\cite{liu_2019_common,liu_2021_risks,makarov_2019_trading,franz_2020_crypto,griffin_2020_is,liu_2022_cryptocurrency}. Our study aims to enhance comprehension of determinants influencing crypto market valuation, emphasizing blockchain transactions and network effects. As juxtaposed in Table \ref{tab:8}, Liu et al. (2019) \cite{liu_2019_common} introduce a three-factor model for cryptocurrency pricing, while Liu and Tsyvinksi (2021) \cite{liu_2021_risks} emphasize the impact of network factors on returns. Liu and Zhang (2022) \cite{liu_2022_cryptocurrency} devise a new price-to-utility (PU) ratio, rooted in classic monetary theory and unique UTXO-based blockchain accounting\cite{liu_2022_deciphering}, to predict bitcoin investment returns. Franz and Vilatin (2020) \cite{franz_2020_crypto} analyze the covered interest rate parity between fiat and cryptocurrency markets, finding that high-frequency trading augments crypto market efficiency. Makarov and Schoar (2019) \cite{makarov_2019_trading} explore arbitrage prospects across cryptocurrencies, while Griffin and Shams (2020) \cite{griffin_2020_is} highlight Tether's role in influencing Bitcoin prices post-market downturns. Our contribution to this discourse delves into transaction and network factor efficiencies, spotlighting the previously overlooked role of decentralization. Additionally, we explore the ramifications of cryptocurrency returns on the decentralization of stablecoins.

\section*{Acknowledgment}

We have benefited from the intellectual conversations at Junior Faculty Seminar Series (JFSS) at Duke Kunshan University, Introduction to DeFi \& Blockchain Research Symposium at Blockchain Accelerator Foundation, MIT Digital Currency Initiatives (DCI) Weekly Meeting at MIT Media Lab, the 3$^{th}$ Blockchain Research in Construction Workshop at Northumbria University Newcastle, the 29$^{th}$ Annual Global Finance Conference featuring 2003 Nobel Prize Laureate in Economics Prof. Robert Engle, CryptoEconDay hosted by CryptoEconLab at Protocol Labs, Blockchain Governance Initiative Network (BGIN Block \#6) at UZH Blockchain Center, and 2022 International Conference on Finance \& Technology (ICFT2022) hosted by Antai College of Economics and Management at Shanghai Jiao Tong University, Crypto Economics Security Conference (CESC) hosted by Berkeley Center for Responsible Decentralized Intelligence (DCI), and Peter Carr Brooklyn Quant Experience (BQE) Seminar Series hosted by Finance and Risk Engineering, NYU Tanton School of Engineering. We Thank Lin Will Cong, Agostino Capponi, and Claudio J. Tessone, Liyi Zhou, Lioba Heimbach, Anton Wahrstatter, Fahad Saleh, Rafael Belchior, Andre Augusto for their insightful comments. Luyao Zhang and Yulin Liu are also with SciEcon CIC, a not-for-profit organization aiming to cultivate interdisciplinary research of profound insights and practical impacts in the United Kingdom. Yulin Liu is also with Shiku Foundation in Switzerland.

\bibliographystyle{ieeetr}
\bibliography{references}
\appendix
\label{appendix}
\footnotesize
\section{Properties of Shannon Entropy} \label{prop}
Shannon entropy\cite{shannon_1948_a,band_2013_quantum,carter_2014_an,gray_2011_entropy} is defined as: 
\[ H(X) = {-\sum_{i=1}^{N}P(v_{i})log_{2}[P(v_{i})]}\\\]
We denote $p_i = Pr(X = x_i)$ and $H_m(p_1,..., p_m) = H(X)$. There are m independent events. 
\begin{enumerate}
    \item $H(p)$ is monotonically decreasing in $p$: an increase in the probability of an event decreases the information from an observed event, and vice versa.
    \item $H(p)$ is monotonically decreasing in p: an increase in the probability of an event decreases the information from an observed event, and vice versa.
    \item $H(p) \ge 0:$ information is a non-negative quantity.
    \item $H(1) = 0:$ events that always occur do not communicate information.
    \item $H(p1, p2) = H(p1) + H(p2):$ the information learned from independent events is the sum of the information learned from each event.
    \item The Shannon entropy$ H(X)$ is a continuous function of $p_i$. If all $p_i$ are equal, $p_i= \frac{1}{m}$, then H is maximal.
\end{enumerate}
Shannon Entropy also follows the following rules: 
\begin{enumerate}
    \item Gibbs inequality: $H(X) \le log(m)$, with equality if and only if $p_i = \frac{1}{m}, i = 1,2,...,m$
    \item $H$ is monotonically increasing with $m$: \\ $H_m(p_1,p_2,...,p_m) \le H_m(\frac{1}{m},\frac{1}{m},...,\frac{1}{m}) \le H_{m+1}(\frac{1}{m+1},\frac{1}{m+1},...,\frac{1}{m+1})$
\end{enumerate}

\section{Meta Data} 
\label{metadata}
\input{tables/table2}
\input{tables/table7}

\section{Additional Figures} \label{addfig}

\begin{figure}[!htbp]
	\includegraphics[width = \linewidth]{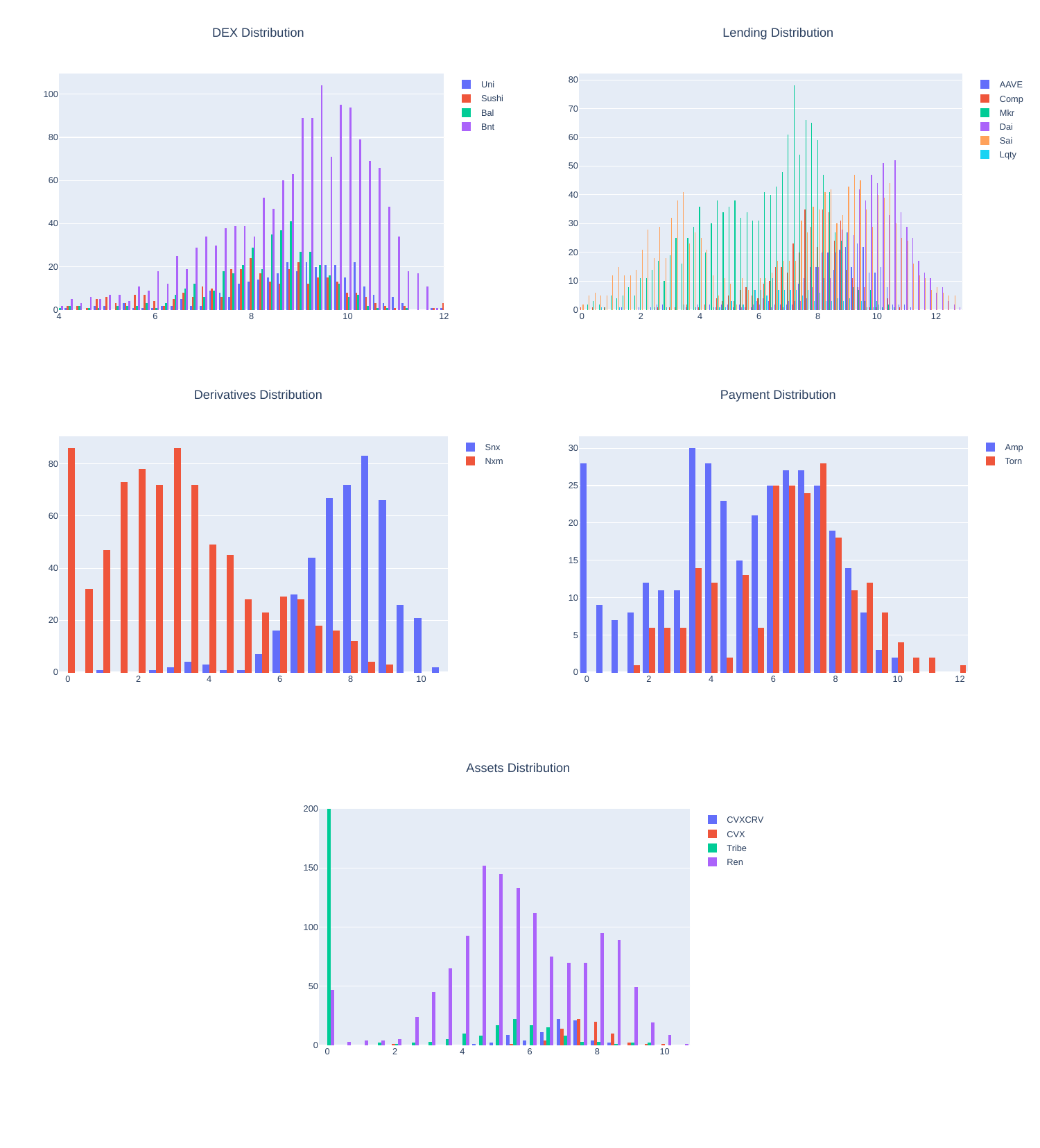}
	\begin{tablenotes}
	\footnotesize
	 \item This figure shows that the level of decentralization of tokens within each category varies greatly: tokens with a higher TVL ranking tend to be more decentralized. This again shows that tokens with greater market demand and utilization are more decentralized.
	\end{tablenotes}
	\caption{Decentralization Index Distributions of Tokens by Category}
	\label{fig:4}
\end{figure}

\begin{figure}[!htbp]
\centering
	\includegraphics[width = 0.6\linewidth]{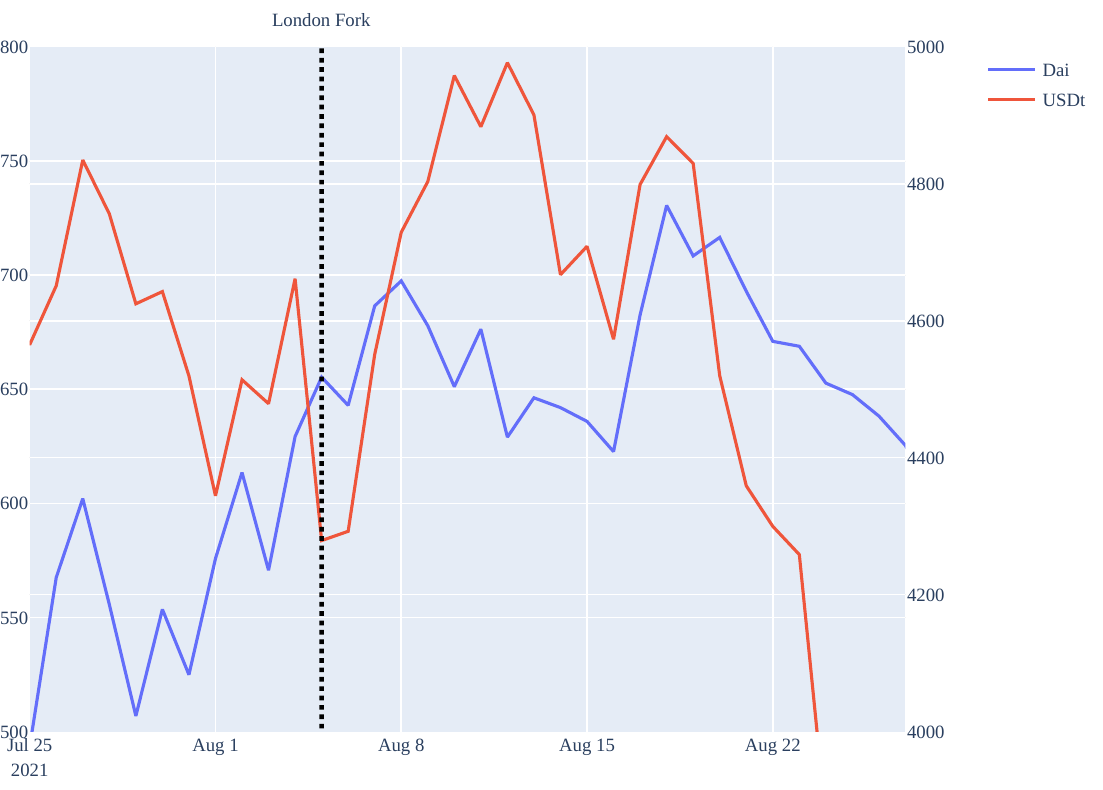}
	\begin{tablenotes}
    \footnotesize
    \item This figure shows the decentralization index of Dai and USDt during the time frame of our EIP 1559 econometric analysis. The figure plots the 30-day simple moving average value of the index. The index values have a daily granularity. The left y-axis is the index value for Dai, the right y-axis is the index value for USDt, and the x-axis is the date.
    \end{tablenotes}
	\caption{EIP-1559: Dai and USDt}
	\label{fig:11}
\end{figure}

\begin{figure}[!htbp]
\centering
 \includegraphics[width = 0.7\linewidth]{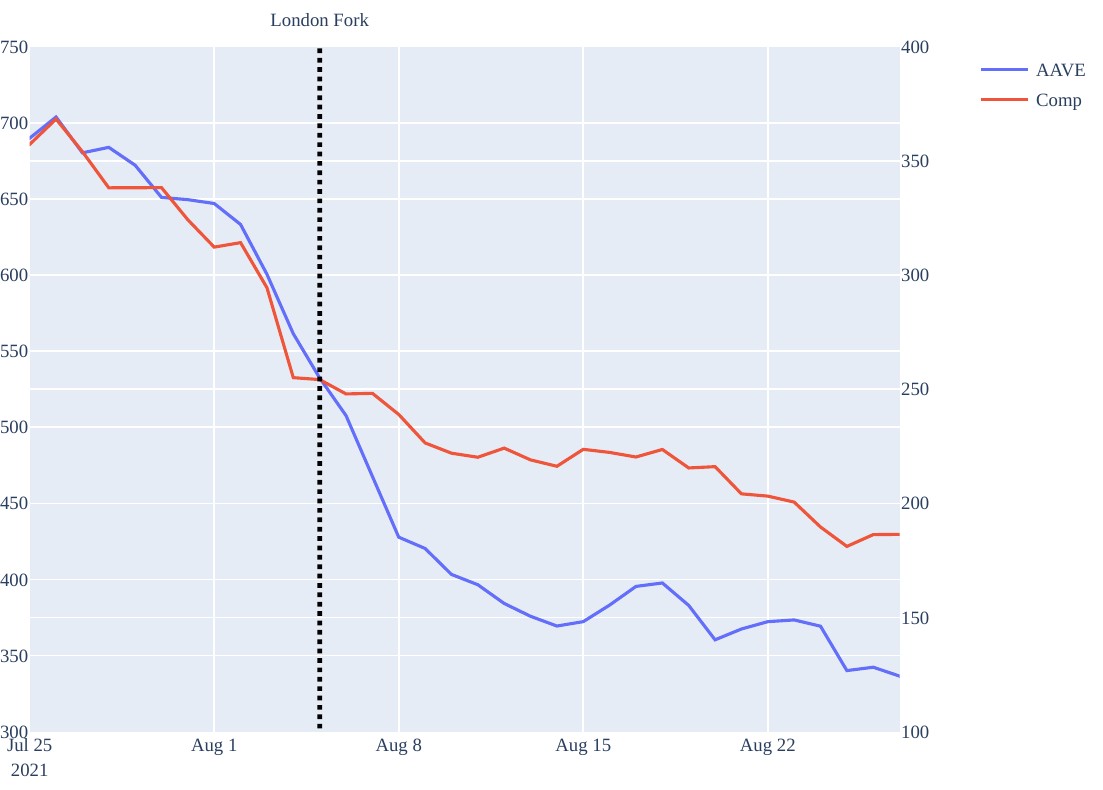}
	\begin{tablenotes}
    \footnotesize
    \item This figure shows the decentralization index of Aave and Compound during the time frame of our EIP-1559 econometric analysis. The figure plots the 30-day simple moving average value of the index. The index values have a daily granularity. The left y-axis is the index value for Aave, the right y-axis is the index value for Compound, and the x-axis is the date.
    \end{tablenotes}
	\caption{EIP-1559: Compound and Aave}
	\label{fig:13}
\end{figure}

\section{Robustness and Sensitivity Checks for Causal 
Analysis}
\label{robustness}

\begin{figure}[!htbp]
\centering
	\includegraphics[width = 0.65\linewidth]{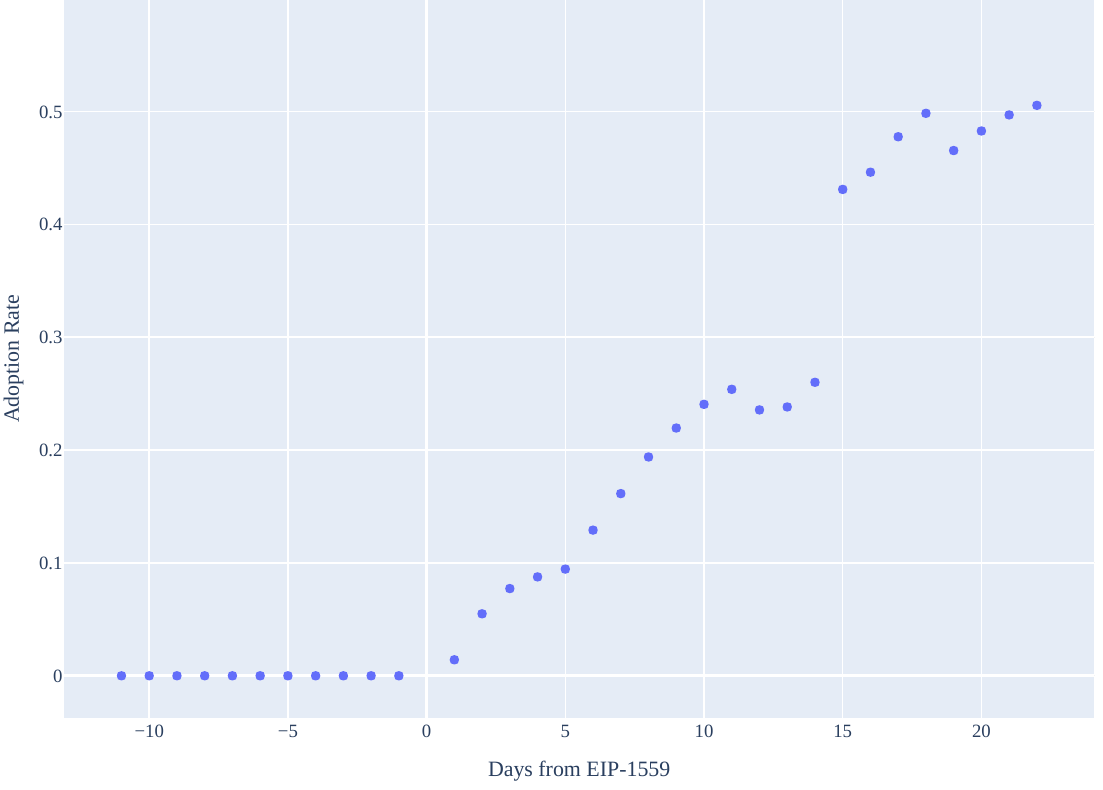}
	\begin{tablenotes}
    \footnotesize
    \item This figure shows the adoption of EIP-1559 during the time frame of our causal analysis. A strong linear relationship exists between the days since EIP-1559 and the adoption rates.
    \end{tablenotes}
	\caption{EIP-1559 Adoption}
	\label{fig:14}
\end{figure}

\begin{figure}[!htbp]
\centering
	\includegraphics[width = 0.65\linewidth]{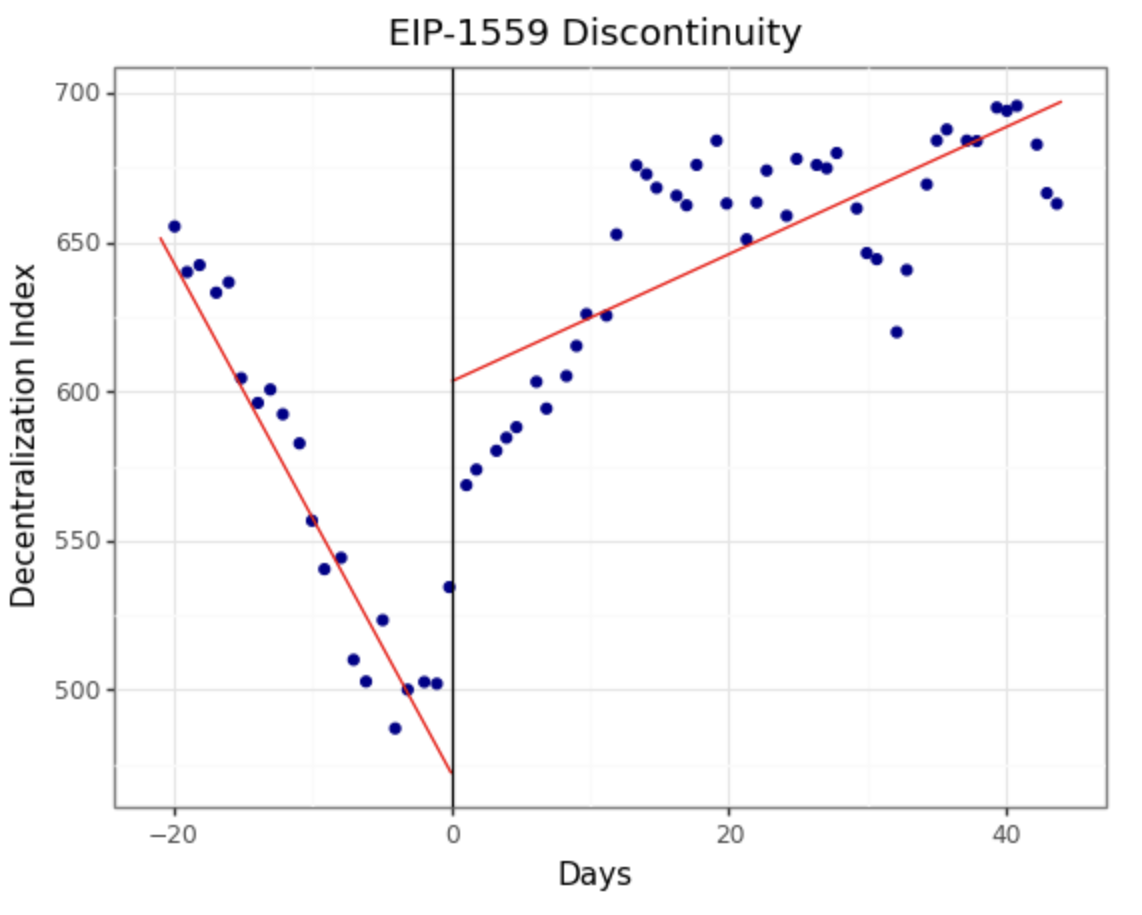}
	\begin{tablenotes}
    \footnotesize
    \item This figure shows the regression discontinuity of Dai with twice the time frame of our EIP-1559 analysis. The figure plots the 30-day simple moving average value of the decentralization index concerning time. The index values have a daily granularity. The y-axis is the index value, and the x-axis is the days from EIP-1559.
    \end{tablenotes}
	\caption{EIP-1559: Discontinuity Twice Bandwidth}
	\label{fig:16}
\end{figure}

\begin{figure}[!htbp]
\centering
	\includegraphics[width = 0.65\linewidth]{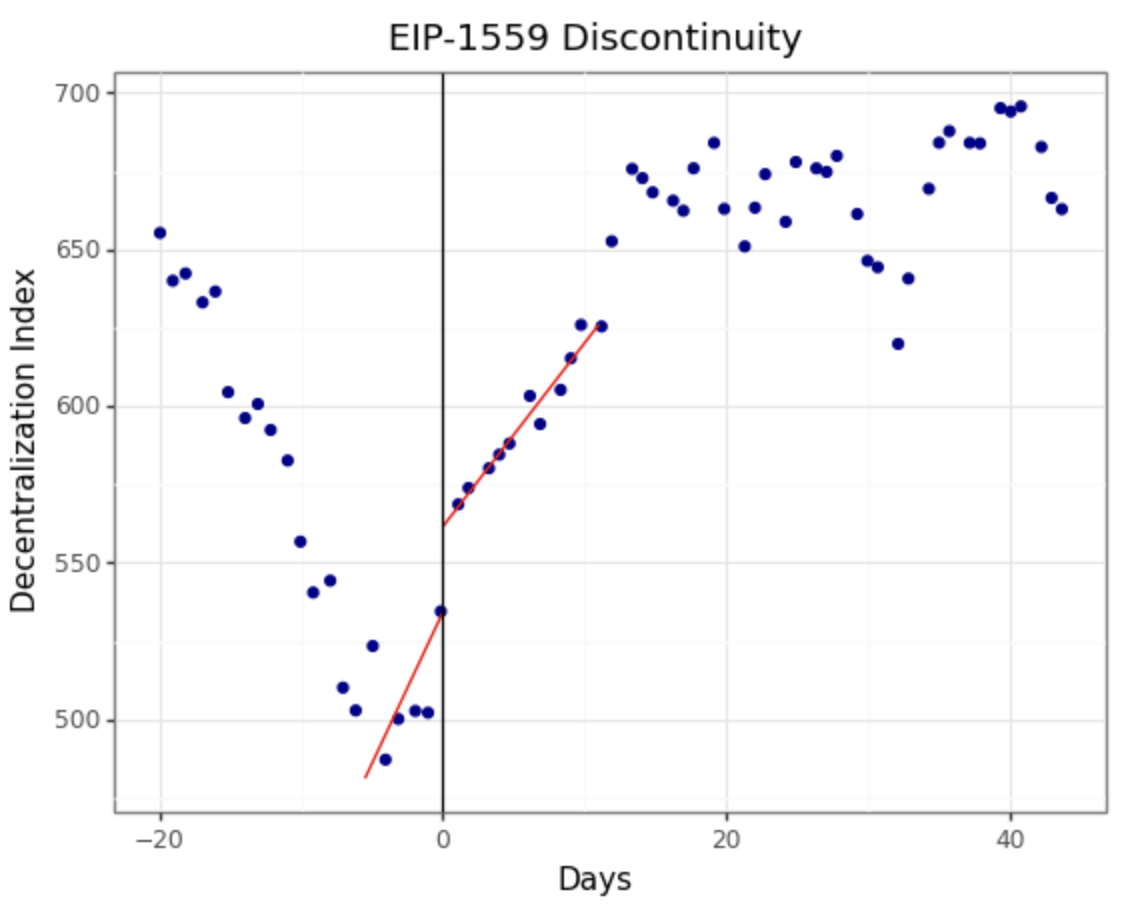}
	\begin{tablenotes}
    \footnotesize
    \item This figure shows the regression discontinuity of Dai within half the time frame of our EIP-1559 analysis. The figure plots the 30-day simple moving average value of the decentralization index concerning time. The index values have a daily granularity. The y-axis is the index value, and the x-axis is the days from EIP-1559.
    \end{tablenotes}
	\caption{EIP-1559: Discontinuity Half Bandwidth}
	\label{fig:17}
\end{figure}

\section{Additional Tables} 
\label{additionalsts}
\input{tables/table9}

%\input{tables/table5}
%\input{tables/table6}
%\input{tables/table3}
%\input{tables/table4}
%\input{tables/table12}
%\input{tables/table13}
%\input{tables/table14}
%\input{tables/table15}
%\input{tables/table19}
\input{tables/table8}

\end{document}

%% file: preamble.tex
\usepackage{cite}
\usepackage{amsmath,amssymb,amsfonts}
\usepackage{algorithmic}
\usepackage{graphicx}
\usepackage{float}
\usepackage{textcomp}
\usepackage{xcolor}
\usepackage{dsfont}
\usepackage{xurl}
\usepackage{threeparttable}
\usepackage{longtable}
\usepackage{subfig}
\usepackage{ragged2e}
\usepackage{MnSymbol}
\usepackage{wasysym} 
\usepackage{fontawesome5}
\usepackage{makecell}
\usepackage{multirow}

\def\BibTeX{{\rm B\kern-.05em{\sc i\kern-.025em b}\kern-.08em
    T\kern-.1667em\lower.7ex\hbox{E}\kern-.125emX}}

\renewcommand{\thefootnote}{\fnsymbol{footnote}}

\newcommand{\symbolfootnote}[1]{\unskip\raisebox{0.6ex}{$\, ^{#1}$}}

\usepackage{pgf}

\newif\ifcomments
\commentstrue

%% file: tables/tab16.tex
\begin{table}[!htbp]
\fontsize{12pt}{12pt}\selectfont
\caption{Blockchain Decentralization Taxonomy}
\label{tab:16}
\resizebox{0.5\textwidth}{!}{
\renewcommand{\arraystretch}{1.0}
\begin{tabular}{|p{.05\textwidth}|p{.06\textwidth}|p{.44\textwidth}|p{.12\textwidth}|p{.14\textwidth}|}
\hline
\textbf{Cite}  & \textbf{Facets} & \textbf{Measures: Notes}& \textbf{Granularity} & \textbf{Assets}  \\ \hline
\hline

\multirow{3}{*}{\cite{cong_2022_inclusion}}      & \( \mathcal{C} \) \( \rightsquigarrow \) & \begin{tabular}[c]{@{}l@{}}-Gini coefficient: \textit{mining power}\\-Shannon entropy: \textit{mining power} \end{tabular}    & \begin{tabular}[c]{@{}l@{}}-\( \mathcal{D} \) \( \sun \)\\ -\( \mathcal{W} \) \( \setminus\mkern-12mu\mid \)\end{tabular} & \multirow{3}{*}{\begin{tabular}[c]{@{}l@{}}ETH \faEthereum\\ ERC20\end{tabular}}                           \\ \cline{2-4} & \( \mathcal{W} \) \( \textdollar \)  & -HHI: \textit{wealth ownership} & -\( \mathcal{D} \) \( \sun \)  & \\ \cline{2-4}
 & \( \mathcal{T} \) \( \rightarrow \)   & -Distribution: \textit{transaction value}   & \begin{tabular}[c]{@{}l@{}}-\( \mathcal{D} \) \( \sun \)\\ -\( \mathcal{W} \) \( \setminus\mkern-12mu\mid \)\end{tabular} &    \\ \hline

\cite{ao_2022_are}  & \( \mathcal{T} \) \( \rightarrow \)  & -Social network features: \newline \textit{transaction value} \newline -Core-peripheral Structure: \newline \textit{transaction network}  & -\( \mathcal{A} \) \( \infty \)  &\begin{tabular}[c]{@{}l@{}} AAVE \end{tabular}\\ \hline

\cite{decollibus_2021_heterogeneous}     & \( \mathcal{T} \) \( \rightarrow \)  & -Preferential attachment: \textit{transaction value} \newline -Probability distribution: \textit{transaction volume}  & -\( \mathcal{A} \) \( \infty \)  & \begin{tabular}[c]{@{}l@{}}ETH \faEthereum\\ LINK\\ USDT\\ BNB\end{tabular} \\ \hline

\cite{campajola_2022_the}& \( \mathcal{C} \) \( \rightsquigarrow \)      & -Nakamoto coefficient:\textit{mining power}  & -\( \mathcal{W} \) \( \setminus\mkern-12mu\mid \) & \begin{tabular}[c]{@{}l@{}}BTC \faBitcoin \\ ETH \faEthereum \\ BCH\\ LTC\\ DOGE\\ MONA\\\end{tabular}\\ 

\cline{2-3} & \( \mathcal{W} \) \( \textdollar \)  & -Gini coefficient: \textit{wealth ownership}      &   &  \\ 

\cline{2-3} & \( \mathcal{T} \) \( \rightarrow \)   & -Clustering of addresses: \textit{transaction network}\newline -Core-periphery structure: \textit{transaction network} & & \\ \hline

\cite{liu_2022_understanding} & \( \mathcal{C} \) \( \rightsquigarrow \)  & -Shannon entropy: \textit{block production} & -\( \mathcal{M} \) \( \Circle \) & EOSIO \\
\hline

\cite{capponi_2021_proofofwork} & \( \mathcal{C} \) \( \rightsquigarrow \)  & -N/A: \textit{game theory analysis} & \( \mathcal{N} \) \( \times \) & N/A \\
\hline

\cite{kwon_2019_impossibility} & \( \mathcal{C} \) \( \rightsquigarrow \) & -Gini coefficient: \textit{block production} & -\( \mathcal{A} \) \( \infty \) & \begin{tabular}[c]{@{}l@{}}BTC \faBitcoin\\ ETH \faEthereum\\ BCH\\ EOS\\ TRON\\ LSK\\ XTZ\\ QTUM\\ WAVES\end{tabular} \\
\hline

\cite{lin_2021_measuring} & \( \mathcal{C} \) \( \rightsquigarrow \)  & -Gini coefficient: \textit{block production}\newline -Shannon entropy: \textit{block production}\newline -Nakamoto coefficient: \textit{block production} & -\( \mathcal{D} \) \( \sun \)\newline -\( \mathcal{W} \) \( \setminus\mkern-12mu\mid \)\newline -\( \mathcal{M} \) \( \Circle \) & \begin{tabular}[c]{@{}l@{}}BTC \faBitcoin\\ ETH \faEthereum\end{tabular} \\
\hline

\cite{kwu_2019_an} & \( \mathcal{W} \) \( \textdollar \) & -Shannon entropy: \textit{address balance} & \multirow{2}{*}{-\( \mathcal{A} \) \( \infty \)} & \begin{tabular}[c]{@{}l@{}}BTC \faBitcoin\\ ETH \faEthereum\end{tabular} \\
\cline{2-3}
& \( \mathcal{C} \) \( \rightsquigarrow \) & -Shannon entropy: \textit{blocks production} & & \\
\hline

\cite{li_2020_comparison} & \( \mathcal{C} \) \( \rightsquigarrow \)  & -Shannon entropy: \textit{block production} & \( \mathcal{N} \) \( \times \) & \begin{tabular}[c]{@{}l@{}}BTC \faBitcoin\\ STEEM\end{tabular} \\
\hline

\cite{li_2020_comparison} & \( \mathcal{C} \) \( \rightsquigarrow \)  & -Shannon entropy: \textit{block production} & \( \mathcal{N} \) \( \times \) & \begin{tabular}[c]{@{}l@{}}BTC \faBitcoin\\ STEEM\end{tabular} \\
\hline

\cite{cong_2020_decentralized} & \( \mathcal{C} \) \( \rightsquigarrow \) & -N/A: \textit{game theory analysis} & -\( \mathcal{A} \) \( \infty \) & BTC \faBitcoin \\
\hline

\cite{chu_2018_the} & \( \mathcal{C} \) \( \rightsquigarrow \) & -N/A: \textit{Byzantine fault tolerance analysis} & \( \mathcal{N} \) \( \times \) & N/A \\
\cline{2-3}
& \( \mathcal{T} \) \( \rightarrow \) & -Top-N address concentration: \newline \textit{transaction value} &  & \\
\hline

\cite{gervais_2014_is} & \( \mathcal{C} \) \( \rightsquigarrow \)  & -N/A: \textit{qualitative analysis} & \( \mathcal{N} \) \( \times \) & BTC \faBitcoin \\
\cline{2-3}
& \( \mathcal{G} \) \( \circlearrowright \) & -N/A: \textit{qualitative analysis} & & \\
\hline

\cite{arnosti_2022_bitcoin} & \( \mathcal{C} \) \( \rightsquigarrow \) & -N/A: \textit{game theory analysis} \newline -HHI: \textit{block production} & \( \mathcal{N} \) \( \times \) & BTC \faBitcoin \\
\hline

\cite{eyal_2018_majority} & \( \mathcal{C} \) \( \rightsquigarrow \) & -N/A: \textit{game theory analysis} & \( \mathcal{N} \) \( \times \)& BTC \faBitcoin \\
\hline

\cite{gencer_2018_decentralization} & \( \mathcal{N} \) \( \circledast \) & -Provisioned bandwidth: \textit{blockchain nodes}\newline -Network structure: \textit{blockchain nodes} & -\( \mathcal{W} \) \( \setminus\mkern-12mu\mid \) & \begin{tabular}[c]{@{}l@{}}BTC \faBitcoin\\ ETH \faEthereum\end{tabular} \\
\cline{2-3}
& \( \mathcal{C} \) \( \rightsquigarrow \)  & -Mining power ratio: \textit{blockchain node} &  & \\
\hline

\cite{srinivasan_2017_quantifying} & \( \mathcal{C} \) \( \rightsquigarrow \) & -Gini coefficient: \textit{mining power} \newline -Nakamoto coefficient: \textit{mining power} & -\( \mathcal{D} \) \( \sun \) \newline -\( \mathcal{A} \) \( \infty \)  & \begin{tabular}[c]{@{}l@{}}BTC \faBitcoin\\ ETH \faEthereum\end{tabular} \\
\cline{2-3}
& \( \mathcal{W} \) \( \textdollar \) & -Gini coefficient: \textit{ownership} \newline -Nakamoto coefficient: \textit{ownership} & & \\
\cline{2-3}
& \( \mathcal{N} \) \( \circledast \) & -Gini coefficient: \newline \textit{node geographical distribution} \newline -Nakamoto coefficient: \newline \textit{node geographical distribution} & & \\
\cline{2-3}
& \( \mathcal{G} \) \( \circlearrowright \) & -Gini coefficient: \newline \textit{developer commits and codebase clients} \newline -Nakamoto coefficient: \newline \textit{developer commits and codebase clients} & & \\
\hline

\cite{lee_2021_dq} & \( \mathcal{N} \) \( \circledast \) & -Gini coefficient: \newline \textit{node geographical distribution} & -\( \mathcal{A} \) \( \infty \) & \begin{tabular}[c]{@{}l@{}}BTC \faBitcoin\\ ETH \faEthereum\end{tabular} \\
\hline

\cite{chen_2020_decentralized} & \( \mathcal{G} \) \( \circlearrowright \) & -Decentralization score: \textit{governance structure} & -\( \mathcal{A} \) \( \infty \) & Unclear \\
\hline

\cite{gu_2020_empirical} & \( \mathcal{G} \) \( \circlearrowright \) & -N/A: \textit{review} & \( \mathcal{N} \) \( \times \) & \begin{tabular}[c]{@{}l@{}}MKR\\ RSR\\ LUNA\\ SC\\ AUG\\ LINK\\ UMA\end{tabular} \\
\hline

\cite{pelt_2020_defining} & \( \mathcal{G} \) \( \circlearrowright \) & N/A: \textit{qualitative analysis} & \( \mathcal{N} \) \( \times \) & \begin{tabular}[c]{@{}l@{}}ETH \faEthereum\\ EOS\end{tabular} \\
\hline

\cite{gupta_2018_gini} & \( \mathcal{W} \) \( \textdollar \) & -Gini coefficient: \textit{address balance} & -\( \mathcal{A} \) \( \infty \) & BTC \faBitcoin \\
\hline

\cite{rocsu2021evolution} & \( \mathcal{W} \) \( \textdollar \) & -N/A: \textit{game theory analysis} & N/A &N/A\\
\hline

\end{tabular}
}

{\raggedright \footnotesize
\vspace{1.0ex}
\fontsize{8pt}{8pt}\selectfont
\emph{Note: This table categorizes recent literature on blockchain decentralization by the facets of blockchain layers, specific measures used, granularities of quantifications, and crypto assets studied. \textbf{Cite}: Citations of the referenced literature   \textbf{Facets}: Different facets of decentralization examined: \( \mathcal{C} \) \( \rightsquigarrow \): Consensus, \( \mathcal{W} \) \( \textdollar \): Wealth, \( \mathcal{T} \) \( \rightarrow \): Transaction, \( \mathcal{G} \) \( \circlearrowright \): Governance, \( \mathcal{N} \) \( \circledast \): Network. \textbf{Measures/Notes}: Specific metrics or qualitative analysis conducted, along with additional notes about the measures. \textbf{Ganularity}: The frequency for which the decentralization metric is measured: \( \mathcal{D} \) \( \sun \) : Daily, \( \mathcal{W} \) \( \setminus\mkern-12mu\mid \): Weekly, \( \mathcal{M} \) \( \Circle \): Monthly, \( \mathcal{A} \) \( \infty \): All---one index for a whole time duration, \( \mathcal{N} \) \( \times \): N/A---no qualification provided. \textbf{Assets}: The short name for blockchain assets studied in each literature: N/A---no specific asset mentioned, short name dictionary: \url{https://coinmarketcap.com/all/views/all/}.}
\par}
\end{table}

%% file: tables/table11.tex
% \usepackage{multirow}
\begin{table}[H]
\caption{Top DeFi Tokens and Categories}
\label{tab:1}
\resizebox{0.5\textwidth}{!}{%
\begin{tabular}{ccccccrrrrc}
\hline
\textbf{Protocol Type}       & \textbf{Rank} & \textbf{Name}                 & \textbf{Tokens} & \textbf{Token Type}  & \textbf{Token Category} & \multicolumn{1}{c}{\textbf{Genesis Date}} & \multicolumn{1}{c}{\textbf{Duration}} & \multicolumn{1}{c}{\textbf{Market Cap}} & \multicolumn{1}{c}{\textbf{TVL}} & \textbf{Data} \\ \hline
\multirow{7}{*}{Lending}     & 1                     & Aave                          & AAVE            & Governance           & Platform                & 2020-10-02                                & 302                                   & 4303                                    & 13440                                           & Yes           \\
                             & 4                     & Instadapp                     & INST            & Governance           & Platform                & 2021-04-06                                & 116                                   & 137                                     & 8840                                            & No            \\
                             & 2                     & Compound                      & COMP            & Governance           & Platform                & 2020-06-14                                & 412                                   & 2198                                    & 9080                                            & Yes           \\
                             & \multirow{2}{*}{5}    & \multirow{2}{*}{Maker}        & MKR             & Governance           & Platform                & 2017-12-15                                & 1324                                  & 2699                                    & \multirow{2}{*}{7690}                           & Yes           \\
                             &                       &                               & DAI/SAI         & Stablecoin           & General Payment         & 2017-12-18                                & 1321                                  & 5400                                    &                                                 & Yes           \\
                             & \multirow{2}{*}{10}   & \multirow{2}{*}{Liquity}      & LQTY            & Dividend Token       & Platform                & 2021-04-05                                & 117                                   & 34                                      & \multirow{2}{*}{1970}                           & Yes           \\
                             &                       &                               & LUSD            & Stablecoin           & General Payment         & 2021-04-15                                & 107                                   & 499                                     &                                                 & Yes            \\ \hline
\multirow{5}{*}{DEX}         & 3                     & Curve                         & CRV             & Governance           & Platform                & 2020-08-12                                & 353                                   & 616                                     & 8990                                            & No            \\
                             & 6                     & Uniswap                       & UNI             & Governance           & Platform                & 2020-09-18                                & 316                                   & 11475                                   & 6160                                            & Yes           \\
                             & 8                     & Sushiswap                     & SUSHI           & Governance           & Platform                & 2020-08-26                                & 339                                   & 1728                                    & 3570                                            & Yes           \\
                             & 12                    & Balancer                      & BAL             & Liquidity            & Platform                & 2020-06-20                                & 406                                   & 240                                     & 1610                                            & Yes           \\
                             & 14                    & Bancor                        & BNT             & Liquidity            & Platform                & 2017-06-17                                & 1505                                  & 857                                     & 1420                                            & Yes           \\ \hline
\multirow{7}{*}{Assets}      & \multirow{2}{*}{7}    & \multirow{2}{*}{Convex}       & CVXCRV          & Yield Farming        & Cash-Flow               & 2021-05-17                                & 75                                    & 119                                     & \multirow{2}{*}{4780}                           & Yes           \\
                             &                       &                               & CVX             & Yield Aggregator     & Platform                & 2021-05-17                                & 75                                    & 51                                      &                                                 & Yes           \\
                             & 9                     & yearn.finance                 & YFI             & Yield Aggregator     & Platform                & 2020-7-17                                 & 379                                   & 1190                                    & 3370                                            & No            \\
                             & \multirow{2}{*}{18}   & \multirow{2}{*}{Fei Protocol} & TRIBE           & Governance           & Platform                & 2020-3-28                                 & 490                                   & 260                                     & \multirow{2}{*}{605}                            & Yes           \\
                             &                       &                               & FEI             & Stablecoin           & General Payment         & 2021-04-03                                & 119                                   & 357                                     &                                                 & No            \\
                             & 20                    & BadgerDAO                     & BADGER          & Yield Aggregator     & Platform                & 2020-11-28                                & 245                                   & 106                                     & 523                                             & No            \\
                             & 19                    & RenVM                         & REN             & Governance           & Platform                & 2017-12-31                                & 1308                                  & 398                                     & 531                                             & Yes           \\ \hline
\multirow{4}{*}{Payment}     & 11                    & Flexa                         & AMP             & Collateral           & Platform                & 2019-01-09                                & 934                                   & 3558                                    & 1660                                            & Yes           \\
                             & 17                    & Tornado Cash                  & TORN            & Governance           & Platform                & 2020-12-18                                & 225                                   & 36                                      & 688                                             & Yes           \\
                             & 28                    & Sablier                       & (no token)      &                      &                         & \multicolumn{1}{c}{}                      & \multicolumn{1}{l}{}                  & \multicolumn{1}{l}{}                    & 316                                             & NA            \\
                             & 38                    & xDai                          & STAKE           & Governance           & Platform                & 2020-04-15                                & 472                                   & 40                                      & 129                                             & No            \\ \hline
\multirow{4}{*}{Derivatives} & 13                    & Synthetix                     & SNX             & Collateral           & Platform                & 2020-05-11                                & 446                                   & 1696                                    & 1560                                            & Yes           \\
                             & 25                    & Nexus Mutual                  & NXM             & Governance           & Platform                & 2019-05-23                                & 800                                   & 684                                     & 416                                             & Yes           \\
                             & 29                    & BarnBridge                    & BOND            & Governance           & Platform                & 2020-08-26                                & 339                                   & 88                                      & 259                                             & No            \\
                             & 32                    & dYdX                          & (no token)      & \multicolumn{1}{l}{} & \multicolumn{1}{l}{}    & \multicolumn{1}{c}{}                      & \multicolumn{1}{l}{}                  & \multicolumn{1}{l}{}                    & 183                                             & NA            \\ \hline
\end{tabular}
}
\vspace{0.5ex}

{\raggedright \footnotesize \emph{Note: This table summarizes key attributes of all the top DeFi tokens ranked (across all DeFi protocols) according to total value locked and indicates whether they are included in our analysis. Token type refers to the conventional industry categorization of tokens according to CoinGecko~\cite{a2021}. Token category refers to the economic usage of tokens described by the Palgrave Handbook of FinTech
and Blockchain ~\cite{cong_2022_inclusion}. Units for Market Cap and TVL are in millions of US dollars.} \par}
\end{table}

%% file: tables/table20.tex
\begin{table}[!htbp]
	\caption{Ether Market Regression on Sai and Dai Decentralization}
	\label{tab:20}
	\begin{minipage}{\columnwidth}
		\begin{center}
			\begin{tabular}{llllll}
				\hline
				\multicolumn{3}{c}{{\small\textit{Dependent variable: Decentralization Index}}}\\[-1.8ex]\\
				\hline
				
				\textbf{Features}   & \thead{\makecell{(1) \\ \textbf{Sai}\\ coefficient\\(std.)}} & \thead{\makecell{(2) \\\textbf{Dai}\\ coefficient\\(std.)}}\\
				\\[-1.8ex]\hline
				ETH\_Ret    &   
				\thead{\makecell{$794.761$\\(1260.203)}}& 	\thead{\makecell{$-1831.393$\\(1366.873)}}\\
		 	    ETH\_Ret7  & \thead{\makecell{$864.508^{*}$\\(520.820)}}& \thead{\makecell{$-102.925$\\(549.664)}}\\
				ETH\_Ret14   & \thead{\makecell{$784.206^{***}$\\(298.941)}}& \thead{\makecell{$169.221$\\(342.763)}}\\
				ETH\_Ret21    & \thead{\makecell{$770.518^{***}$\\(226.084)}}& \thead{\makecell{$391.680$\\(266.600)}}\\
				ETH\_Ret30   & \thead{\makecell{$813.912^{***}$\\(182.544)}}& \thead{\makecell{$443.710^{**}$\\(191.323)}}\\
				ETH\_VtyDayRet30d      & \thead{\makecell{$-18345.301^{***}$\\(3138.375)}}& \thead{\makecell{$-32.986$\\(2136.514)}}\\
			    ETH\_PC     & 
			    \thead{\makecell{$536.188^{***} $\\(137.019)}}& 
			    \thead{\makecell{$254.369$\\(158.502)}}\\
				\hline
			\end{tabular}
		\end{center}
            \footnotesize  $^{*}$p$<$0.1; $^{**}$p$<$0.05; $^{***}$p$<$0.01\\ 
		 We concluded that all variables are stationary using the augmented Dickey-Fuller test for stationarity. \\
		 The Newy-West estimator for standard errors was used to adjust for high autocorrelation with day $t-1$. Column (1) shows the results for Sai, and column (2) shows the results for Dai.
	\end{minipage}
\end{table}%  

%% file: tables/table18.tex
\begin{table}[!htbp] \centering
  \caption{EIP-1559 Regression Discontinuity (Dai)}
  \label{tab:18}
  \begin{minipage}{0.48\textwidth}
		\begin{center}
\begin{tabular}{p{0.2\textwidth}*{4}{>{\centering\arraybackslash}p{0.2\textwidth}}}
\hline
& \multicolumn{3}{c}{\textit{Dependent variable: Decentralization Index SMA30}}\\[-1.8ex] \\ 
\cline{2-4}
\
\\[-1.8ex] & (1) & (2) & (3) \\
\\ [-1.8ex]
\hline \\[-1.8ex]
 EIP & 73.427$^{***}$ & 69.990$^{***}$ & 42.241$^{***}$ \\
  & (13.610) & (12.138) & (10.733) \\
 Day & -3.865$^{**}$ & -3.679$^{**}$ & -2.607$^{**}$ \\
  & (1.657) & (1.469) & (1.130) \\
 EIP\_Day & 9.339$^{***}$ & 9.603$^{***}$ & 4.951$^{***}$ \\
  & (1.757) & (1.550) & (1.513) \\
 TxTfrValAdjUSD & & -0.000$^{}$ & -0.000$^{*}$ \\
  & & (0.000) & (0.000) \\
 TxTfrCnt & & 0.004$^{**}$ & 0.001$^{}$ \\
  & & (0.001) & (0.001) \\
 ROI & & & 5239.451$^{}$ \\
  & & & (5137.365) \\
 VtyDayRet30d & & & 510364.068$^{***}$ \\
  & & & (104645.070) \\
 Intercept & 495.379$^{***}$ & 447.263$^{***}$ & 340.328$^{***}$ \\
  & (11.241) & (24.662) & (28.603) \\
  \\
\hline \\[-1.8ex]
 Observations & 33 & 33 & 33 \\
 $R^2$ & 0.933 & 0.952 & 0.975 \\
 F Statistic & 134.707$^{***}$  & 106.272$^{***}$ & 140.968$^{***}$  \\
\hline
\textit{Note:} & \multicolumn{3}{r}{$^{*}$p$<$0.1; $^{**}$p$<$0.05; $^{***}$p$<$0.01} \\
\end{tabular}
\end{center}

	\footnotesize This table displays the linear regression results on the 30-day simple moving average of our decentralization index for the stablecoin Dai.  EIP is an indicator variable for the London Hardfork on August 5, 2021. Day is the number of days from the London Hardfork. EIP\_Day is the interaction variable of EIP and Day. TxTfrValAdjUSD is the daily transaction volume in USD of Dai. TxTfrCnt is the daily transaction count of Dai. ROI is the 1-day return on investment. VtyDayRet30d is the 30-day volatility of Dai. \\

\end{minipage}
\end{table}

%% file: tables/table2.tex
% \usepackage{graphicx}
\begin{table}[!htbp]
\centering
\caption{Meta Data}
\resizebox{.45\textwidth}{!}{%
\begin{tabular}{|l|l|l|l|l|}
\hline

{\color[HTML]{000000} \textbf{Variable Name}} & {\color[HTML]{000000} \textbf{Type}} & {\color[HTML]{000000} \textbf{Unit}} & {\color[HTML]{000000} \textbf{Range}} & {\color[HTML]{000000} \textbf{Description}} \\ \hline
{\color[HTML]{000000} value} & {\color[HTML]{000000} Numeric} & {\color[HTML]{000000} wei} & {\color[HTML]{000000} {[}$0,+\infty$)} & {\color[HTML]{000000} Value of transaction in Wei} \\ \hline
{\color[HTML]{000000} from\_address} & {\color[HTML]{000000} String} & {\color[HTML]{000000} /} & {\color[HTML]{000000} /} & {\color[HTML]{000000} Address of the sender} \\ \hline
{\color[HTML]{000000} to\_address} & {\color[HTML]{000000} String} & {\color[HTML]{000000} /} & {\color[HTML]{000000} /} & {\color[HTML]{000000} Address of the receiver} \\ \hline
{\color[HTML]{000000} block\_timestamp} & {\color[HTML]{000000} Timestamp} & {\color[HTML]{000000} /} & {\color[HTML]{000000} {[}Token Genesis to 2021-07-31{]}} & {\color[HTML]{000000} Time of the transaction} \\ \hline
{\color[HTML]{000000} token\_address} & {\color[HTML]{000000} String} & {\color[HTML]{000000} /} & {\color[HTML]{000000} /} & {\color[HTML]{000000} Name of ERC-20 token} \\ \hline
{\color[HTML]{000000} CapMrktCurUSD} & {\color[HTML]{000000} Numeric} & {\color[HTML]{000000} USD} & {\color[HTML]{000000} {[}$0,+\infty$)} & {\color[HTML]{000000} Market capitalization} \\ \hline
{\color[HTML]{000000} PriceUSD} & {\color[HTML]{000000} Numeric} & {\color[HTML]{000000} USD} & {\color[HTML]{000000} {[}$0,+\infty$)} & {\color[HTML]{000000} Price in USD} \\ \hline
{\color[HTML]{000000} VtyDayRet30d} & {\color[HTML]{000000} Numeric} & {\color[HTML]{000000} /} & {\color[HTML]{000000} {[}$0,+\infty$)} & {\color[HTML]{000000} The 30 day volatility} \\ \hline
{\color[HTML]{000000} TxTfrValAdjUSD} & {\color[HTML]{000000} Numeric} & {\color[HTML]{000000} USD} & {\color[HTML]{000000} {[}$0,+\infty$)} & {\color[HTML]{000000} Daily transaction volume} \\ \hline
{\color[HTML]{000000} TxTfrCnt} & {\color[HTML]{000000} Numeric} & {\color[HTML]{000000} /} & {\color[HTML]{000000} {[}$0,+\infty)$} & {\color[HTML]{000000} Daily transaction count} \\ \hline
\end{tabular}%
}
\vspace{0.5ex}

{\raggedright \footnotesize \emph{Note: This table displays the meta data we acquired from Google BigQuery and CoinMetrics.} \par}

\label{tab:2}
\end{table}

%% file: tables/table7.tex
% \usepackage{graphicx}
\begin{table}[!htbp]
\caption{Data Dictionary}
\label{tab:7}
\resizebox{0.5\textwidth}{!}{%
\begin{tabular}{l|ll}
\textbf{Variable}       & \textbf{Description}                                                                                                                &  \\ \hline
\textbf{val}            & Value of decentralization index                                                                                                     &  \\
\textbf{date}           & Date of index and market variables                                                                                                  &  \\
\textbf{CapMrktCurUSD}  & Market capitalization in USD: The sum USD value of the current supply.                                                              &  \\
\textbf{PriceUSD}       & The fixed closing price of the asset as of 00:00 UTC the following day                                                              &  \\
\textbf{VtyDayRet30d}   & The 30 day volatility, measured as the standard deviation of the natural log of daily returns over the past 30 days.                &  \\
\textbf{TxTfrValAdjUSD} & The USD value of the sum of native units transferred between distinct addresses that interval removing noise and certain artifacts. &  \\
\textbf{TxTfrCnt}       & Daily transaction count                                                                                                             &  \\
\textbf{ETH\_Ret}       & 1 day return of Ether                                                                                                               &  \\
\textbf{ETH\_Ret7}      & 7 day return of Ether                                                                                                               &  \\
\textbf{ETH\_Ret14}     & 14 day return of Ether                                                                                                              &  \\
\textbf{ETH\_Ret21}     & 21 day return of Ether                                                                                                              &  \\
\textbf{ETH\_Ret30}     & 30 day return of Ether                                                                                                              &  \\
\textbf{ETH\_PC}        & First principle component of all Ether returns: Ret to Ret\_30                                                                      &  \\
\textbf{ROI}            & 1 day return of asset                                                                                                               &  \\
\textbf{EIP}            & Indicator variable for the London Hardfork on August 5, 2021                                                                        &  \\
\textbf{Day}            & Number of days from the London Hardfork                                                                                             &  \\
\textbf{EIP\_Day}       & Interaction variable of EIP and Day                                                                                                 & 
\end{tabular}
}

\vspace{0.5ex}
{\raggedright \footnotesize \emph{Note: This table is the data dictionary describing all the decentralization index and economic variables we use in the empirical results section.} \par}
\end{table}

%% file: tables/table9.tex
% \usepackage{graphicx}
\begin{table}[!htbp]
\caption{Sai and Dai Events}
\label{tab:9}
\resizebox{0.5\textwidth}{!}{%
\begin{tabular}{p{3cm}lp{4cm}l}
\textbf{Event} & \textbf{Date} & \textbf{Description} & \textbf{Source} \\ \hline
Dai Genesis & Dec. 19, 2019 & Multicollateral Dai stablecoin created to replace the single-collateral Sai & Dai White Paper \\ \hline
Bitcoin Crash & Mar. 13, 2020 & Bitcoin loses 55\% of market value & Forbes \\ \hline
Transfer Governance Token to MKR & Mar. 25, 2020 & Transfer of MKR Token control to MKR holders (governance community) & Maker Blog \\ \hline
Reintroduce Stability Fees & Sep. 15, 2020 & Maker community voted to reintroduce stability fees & Crypto Briefing \\ \hline
Foundation Returns all Funds to DAO & May 3, 2020 & MakerDAO foundation returns all development funds to the MakerDAO & Maker Blog \\ \hline
MakerDAO Fully Decentralized & July 20, 2021 & MakerDAO completely decentralized after passing core units for self-sufficient operation & Maker Blog
\end{tabular}%
}

\vspace{0.5ex}
{\raggedright \footnotesize \emph{Note: This table describes significant governance events in the history of Sai and Dai, along with market events that are associated with large changes in decentralization.} \par}
\end{table}

%% file: tables/table8.tex
% \usepackage{graphicx}

\begin{table}[!htbp]
\caption{Literature on Crypto-Asset Pricing and Returns}
\label{tab:8}
\resizebox{0.5\textwidth}{!}{%
\begin{tabular}{lllll}
\textbf{Paper} & \textbf{Factors} & \textbf{Economic Variables} & \textbf{Assets} & \textbf{Outcomes} \\ \hline
\begin{tabular}[c]{@{}l@{}}~\cite{liu_2021_risks} \end{tabular} & \begin{tabular}[c]{@{}l@{}}Production, network (wallet \\ user growth, active address \\ growth, transaction count growth, \\ and payment count growth), \\ computing, electricity, currency\end{tabular} & Returns & \begin{tabular}[c]{@{}l@{}}All coins with \\ market capitalization\\  greater than 1,000,000 \\ USD.\end{tabular} & \begin{tabular}[c]{@{}l@{}}Cryptocurrency returns strongly\\  respond to cryptocurrency \\ network factors\end{tabular} \\ \hline
\begin{tabular}[c]{@{}l@{}}\cite{liu_2019_common} \end{tabular} & \begin{tabular}[c]{@{}l@{}}Cryptocurrency market, size, \\ and momentum\end{tabular} & Returns & \begin{tabular}[c]{@{}l@{}}All coins with \\ market capitalization \\ greater than 1,000,000 \\ USD\end{tabular} & \begin{tabular}[c]{@{}l@{}}Three factor model in \\ cryptocurrency pricing\end{tabular} \\ \hline
\begin{tabular}[c]{@{}l@{}}\cite{franz_2020_crypto}\end{tabular} & \begin{tabular}[c]{@{}l@{}}Liquidity, volatility, and bid-ask \\ spreads\end{tabular} & CIP Deviations & Bitcoin (BTC), USD & \begin{tabular}[c]{@{}l@{}}Decrease in covered interest \\ rate parity (CIP) after Q1 2018 \\ due to high-frequency trading\end{tabular} \\ \hline
\begin{tabular}[c]{@{}l@{}}\cite{makarov_2019_trading}\end{tabular} & \begin{tabular}[c]{@{}l@{}}Price deviations across \\ cryptocurrencies, exchanges, \\ and countries\end{tabular} & Arbitrage & \begin{tabular}[c]{@{}l@{}}Bitcoin (BTC), Ethereum \\ (ETH), and Ripple (XRP)\end{tabular} & \begin{tabular}[c]{@{}l@{}}There are arbitrage opportunities \\ across exchanges; Decomposed \\ common and idiosyncratic \\ components of bitcoin returns\end{tabular} \\ \hline
\begin{tabular}[c]{@{}l@{}}\cite{liu_2020_do}\end{tabular} & \begin{tabular}[c]{@{}l@{}}Technological sophistication,\\  ICO success,  liquidity measure\\ and the delisting probability\\ measure\end{tabular} & Returns & \begin{tabular}[c]{@{}l@{}}All ICOs from \\ trackico.com\end{tabular} & \begin{tabular}[c]{@{}l@{}}Technological sophistication is \\ an important determinant of \\ cryptocurrency valuations\end{tabular}\\
\hline
\begin{tabular}[c]{@{}l@{}}\cite{liu_2022_cryptocurrency} \\\end{tabular} & \begin{tabular}[c]{@{}l@{}} Price to utility (PU) ratio,\\ PE ratio (EPR), inverse \\of the NVT ratio (TVN), \\inverse of the P/M ratio\\ (MPR), and inverse of the\\ PU ratio (UPR) \end{tabular} & Returns & \begin{tabular}[c]{@{}l@{}}Bitcoin (BTC)\end{tabular} & \begin{tabular}[c]{@{}l@{}} PU ratio may be effectively\\ used to long-term returns and \\bull markets\end{tabular}\\
\hline
\end{tabular}}

\vspace{0.5ex}

{\raggedright \footnotesize \emph{Note: This table summarizes the factors, economic variables, and assets used from the crypto-asset valuation literature.} \par}
\end{table}